\documentclass[]{interact}

\usepackage{epstopdf}
\usepackage[caption=false]{subfig}

\usepackage[numbers,sort&compress]{natbib}
\bibpunct[, ]{[}{]}{,}{n}{,}{,}

\theoremstyle{plain}
\newtheorem{theorem}{Theorem}[section]

\theoremstyle{definition}

\theoremstyle{remark}

\usepackage{mathtools}
\usepackage{multirow}
\usepackage[strings]{underscore}

\newcommand{\E}{\mathop{\mathbb{E}}}
\newcommand{\undertext}[2]{\ensuremath{\underset{#2}{\underbrace{#1}}}}

\DeclareMathOperator{\tr}{tr}

\DeclareMathOperator{\logit}{logit}

\begin{document}

\articletype{Original Research Article}

\title{Joint Modeling of An Outcome Variable and Integrated Omic Datasets Using GLM-PO2PLS}

\author{
\name{Zhujie~Gu\textsuperscript{a}\thanks{CONTACT Zhujie Gu, Email: z.gu@umcutrecht.nl},
	 Said~el~Bouhaddani\textsuperscript{a}, Jeanine~Houwing-Duistermaat\textsuperscript{b,a,c}, and Hae-Won~Uh\textsuperscript{a}}
\affil{\textsuperscript{a}Dept. of Data Science and Biostatistics, Julius Centre, UMC Utrecht, The Netherlands; \\ \textsuperscript{b}Dept. of Statistics, University of Leeds, The United Kingdom; \\
\textsuperscript{c}Dept. of Statistical Sciences  "Paolo Fortunati" , University of Bologna, Italy.}
}

\maketitle

\begin{abstract}
	In many studies of human diseases, multiple omic datasets are measured. Typically, these omic datasets are studied one by one with the disease, thus the relationship between omics is overlooked.  
Modeling the joint part of multiple omics and its association to the outcome disease will provide insights into the complex molecular base of the disease.
In this article, we extend dimension reduction methods which model the joint part of omics to a novel method that jointly models an outcome variable with omics.
We establish the model identifiability and develop EM algorithms to obtain maximum likelihood estimators of the parameters for normally and Bernoulli distributed outcomes.  Test statistics are proposed to infer the association between the outcome and omics, and their asymptotic distributions are derived.
Extensive simulation studies are conducted to evaluate the proposed model. The method is illustrated by modeling Down syndrome as outcome and methylation and glycomics as omics datasets. Here we show that our model provides more insight by jointly considering methylation and glycomics.
\end{abstract}

\begin{keywords}
Dimension reduction; PLS methods; Multiple omics; Generalized linear models; Data integration.
\end{keywords}

\section{Introduction}
The biological mechanisms underlying human diseases are often complex. Diverse omic datasets represent various aspects of these mechanisms. Recent advances in high-throughput technologies have made it affordable to measure these omic levels for many studies. Typically, these datasets are studied one-by-one. A good example is the analysis of genomic data in more than 5700 GWAS conducted to identify the genetic risk variants associated with more than 3000 traits and human diseases~\cite{Watanabe2019,Uffelmann2021}. 
Other examples include studies of methylation data to pinpoint differentially methylated regions of DNA as indicators of many diseases~\cite{Sheikhpour2021,Nishiyama2021}, and studies of glycomic data to gain insight into the role of post-translational modification of protein in disease pathways~\cite{Tabang2021,Sugar2021}. Though these studies on a single omic dataset provided biological insights of diseases from various aspects, they ignored the correlations among the omic levels. Analyzing multiple linked omic datasets jointly can bring further insights into the biological system underlying diseases. In this paper, we propose a new model for two omic datasets and an outcome variable, where the relationship of the omic datasets with the outcome is modeled via the joint parts of the omic datasets.

Our motivating dataset comes from a family-based case-control study of Down syndrome (DS). DS is the most frequent genomic aneuploidy with an incidence of approximately 1 in 700 live-newborn~\cite{Patterson2007}, caused by the trisomy of all or part of chromosome 21 (trisomy 21). Studies at the molecular level of DS have reported several alterations in methylation~\cite{Franceschi2019,Gensous2020,Bacalini2015,Ciccarone2018} and glycomics~\cite{Franceschi2019,Borelli2015,Cindric2021}. These alternations are mainly discovered by testing the mean difference of a single CpG site or glycan between the DS subjects and healthy controls. Furthermore, these studies were conducted on each omic level separately, overlooking the influence of methylation on glycosylation~\cite{Wahl2018}. We will use our new model to jointly analyze DS and both omics, aiming to investigate whether the molecules involved in the relationship between methylation and glycomics are related to DS.

Generalized linear models (GLM) are flexible models which link linear predictors with an outcome variable via functions such as identity and logit~\cite{Nelder1972a}. 
However, omic datasets are often high-dimensional ($p>>N$) and the features are highly correlated, which leads to (near) collinearity, hence  GLM cannot directly be employed. A solution is to use its penalized variants such as ridge regression~\cite{Hoerl2000} and elastic net~\cite{Zou2005}, which handle the dimensionality and correlation better. To incorporate information from more than one dataset, stacked sets of omic features can be modeled. However, such approaches do not model the relationship between omics, and hence cannot provide insight on the joint part. Furthermore, when the omic datasets are heterogeneous, regressing on a stacked dataset can lead to inferior performance than using only one of the available omic sources~\cite{Rodriguez-Girondo2018,Tissier2018b}.

To model the joint part of two omic datasets, several dimension reduction methods have been developed, which map both datasets from the original high-dimensional spaces to low-dimensional spaces including a joint space which retains the relationship. Among these, PLS~\cite{WOLD1973, TieJong1993} simultaneously decomposes two datasets $ x $ and $ y $ into joint and residual subspaces. The joint (low-dimensional) subspace of one dataset represents the best approximation of $x$ or $y$. The joint subspaces of PLS may contain omic-specific variation due to the presence of heterogeneity between the omics. To separate this omic-specific variation from the joint subspaces, two-way orthogonal partial least squares (O2PLS)~\cite{Trygg2003, Bouhaddani2016a} was proposed which decomposes each dataset into joint, data-specific, and residual subspaces. The data-specific subspaces in $x$ and $y$ capture variation unrelated to each other, and improve the estimates of the joint subspaces for the true relationship between $ x $ and $ y $. A drawback of PLS and O2PLS is that they are algorithmic, and hence do not model the whole distribution. To enable statistical inference, likelihood-based probabilistic approaches such as supervised integrated factor analysis (SIFA)~\cite{Li2017} and probabilistic PLS (PPLS)~\cite{ElBouhaddani2018} were proposed. Recently, a general framework probabilistic O2PLS (PO2PLS)~\cite{Bouhaddani2021} was proposed which contains SIFA and PPLS as specific cases. In PO2PLS, the relationship between two omics is inferred by using a Wald-type test-statistic to test the hypothesis that the joint latent variables of the two omics are related. However, PO2PLS is unsupervised, i.e., the the outcome variable is not used in the joint dimension reduction.
In this paper, we will propose a new model GLM-PO2PLS which extends PO2PLS by including an outcome variable in the model next to the omic datasets. 
The relationship between two omic data is modeled by joint and omic-specific latent variables to deal with possible heterogeneity. The joint latent variables are linked to an outcome variable by a generalized linear model. We develop EM algorithms to obtain maximum likelihood estimators of the parameters for normally and Bernoulli distributed outcomes. The relationship between the outcome variable and the omics and that between two omic datasets can be inferred. The code is available on GitHub (github.com/zhujiegu/GLM-PO2PLS).

The rest of the paper is organized as follows. In Section 2, the PO2PLS model is recapped, and the GLM-PO2PLS model is formulated.
The EM algorithms to estimate its parameters are proposed. Also, two chi-square tests of the relationship between outcome and both omics are proposed. In Section 3, the performance of GLM-PO2PLS is studied for a range of simulation scenarios where the focus is on parameter estimation and outcome prediction performance. In Section 4, we apply GLM-PO2PLS to the motivating DS datasets. We conclude with a discussion.

	\section{Methods}

The GLM-PO2PLS was developed based on PO2PLS model which has been described in detail elsewhere~\cite{Bouhaddani2021}. Briefly, let $x$ and $y$ be two random row-vectors of dimensions $p$ and $q$, respectively. In PO2PLS, $x$ and $y$ are decomposed into joint ($t$ and $u$ of size $r$), specific ($t_\perp$ and $u_\perp$ of size $r_x$ resp. $r_y$) and residual ($e$ and $f$ of size $p$ resp. $q$) parts. Heterogeneity between the joint parts is represented by an additional random vector $h$. The PO2PLS model is written as
\begin{flalign*}
	\begin{aligned}
		x = tW^\top + t_\perp W_\perp^\top + e, \quad
		y = uC^\top + u_\perp C_\perp^\top + f, \quad
		u = tB + h, 
	\end{aligned}
\end{flalign*}
where $ W \, (p \times r) $ and $ C \, (q \times r) $ are the loading matrices for the joint spaces of $x$ and $y$ respectively and  $W_\perp \, (p \times r_x)$ and $C_\perp \, (q \times r_y)$ are the loading matrices for the specific parts of $x$ and $y$ respectively. The $r \times r$ diagonal matrix $B$ models the relationship between the joint components $t$ and $u$. 
With regard to the random vectors, we assume that $t$, $t_\perp$, $u_\perp$, $h$ are zero mean multivariate normally distributed, with diagonal covariance matrices $\Sigma_t$, $\Sigma_{t_\perp}$, $\Sigma_{u_\perp}$, $\Sigma_h$, respectively. Since $ u = tB+h $, the covariance matrix of $ u $ is $ \Sigma_u = B^\top \Sigma_t B + \Sigma_h $. The residual random vectors $e$, and $f$ are independent normally distributed, with zero mean and respective diagonal covariance matrices , $\sigma_e^2 I_p$, and $\sigma_f^2 I_q$, where $ I_p $ and $ I_q $ are identity matrices of size $ p $ and $ q $.

\subsection{The GLM-PO2PLS model}
GLM-PO2PLS jointly models an outcome variable $ z $ with two omic datasets $x$ and $y$, where it is assumed that the effect of $x$ and of $y$ on $z$ is solely through the joint parts of $x$ and $y$.

Using the same notations as in the PO2PLS model, the GLM-PO2PLS model is given by 
\begin{flalign}
	\begin{aligned} \label{For:model}
		x = tW^\top + t_\perp W_\perp^\top + e, \quad
		y & = uC^\top + u_\perp C_\perp^\top + f, \quad
		u = tB + h, \\
		\eta(\mathbb{E}[z]) & = a_0 + ta^\top + ub^\top,
	\end{aligned}
\end{flalign}
with $ a_0 $  the intercept, $ a $ and $ b $  both row-vectors of size $ r $ and and $\eta$ the link function which links the outcome $z$ to the linear predictor $ a_0 + ta^\top + ub^\top$. Note that the equations in the first row of~\eqref{For:model} are identical to the PO2PLS model. Since the joint latent variables $t$ and $u$ are linked to $x$, $y$, and $z$, GLM-PO2PLS jointly models the outcome and two omics.

Now, $u$ is a linear function of $t$, namely  $ u=tB+h $. Hence, the model for $z$ in~\eqref{For:model} can equivalently be written in terms of $ t $ and the $ h $ (the part in $ u $ independent of $ t $), i.e.
\begin{equation}\label{eqab}
	\eta(\mathbb{E}[z]) = a_0 + ta^\top + (tB+h)b^\top = a_0 + t\tilde{a}^\top + h\tilde{b}^\top,
\end{equation}
where $\tilde{a} = a + Bb^\top$ and $\tilde{b} = b$.
With this equivalent parametrization, instability due to near colinearity in the linear predictor of $z$ is reduced. 

In the remainder of the paper, we use the rightmost form in~\eqref{eqab} and omit the tildes on $a$ and $b$. 

\subsection{The GLM-PO2PLS model with a normally distributed outcome}		
In this subsection, we first consider a continuous outcome $z$. The details for a binary $ z $ is then given in the next subsection.  As link function, we use the identity, $\eta(v) = v$. We assume that the outcome is centered and since $t$ and $h$ have zero-mean,  the intercept $ a_0 $ can be omitted.
We assume that the residual $g$ ($g=z - ta^\top - ub^\top$) is normally distributed, $g \sim \mathcal{N}(0, \sigma_g^2)$. 
Since $ (x,y,z) $ is linearly dependent on $ (t,u,t_\perp, u_\perp, e,f,h,g) $, it follows a multivariate normal distribution $ \mathcal{N}(0,\Sigma_\theta) $, with a covariance matrix given by
\begin{equation} \label{For:sigtheta}
	\Sigma_\theta=
	\begin{bmatrix}
		W\Sigma_t W^\top + W_\perp \Sigma_{t_\perp} W_\perp^\top + \sigma_e^2 I_p	&	W \Sigma_t B C^\top	&	W \Sigma_t a^\top\\
		C B \Sigma_t W^\top	&	C \Sigma_u C^\top + C_\perp \Sigma_{u_\perp} C_\perp^\top + \sigma_f^2 I_q	&	C (\Sigma_h b^\top + B \Sigma_t a ^\top)\\
		a \Sigma_t W^\top	&	(a\Sigma_t B + b\Sigma_h)C^\top	&	a\Sigma_t a^\top + b\Sigma_h b^\top + \sigma_g^2
	\end{bmatrix},
\end{equation}
where $ \theta = \{W,C,W_\perp, C_\perp, a, b, B, \Sigma_t, \Sigma_{t_\perp}, \allowbreak \Sigma_{u_\perp}, \sigma_e^2, \sigma_f^2, \Sigma_h, \sigma_g^2\} $ is the collection of GLM-PO2PLS model parameters.

\paragraph*{Identifiability of GLM-PO2PLS}
Latent variable models are typically unidentifiable due to rotation indeterminacy of the loading components. In PO2PLS, identifiability up to sign has been shown under mild conditions~\cite{Bouhaddani2021}. Namely, the loading matrices are semi-orthogonal, i.e. $ W^\top W = C^\top C = I_r $, $ W_\perp^\top W_\perp = I_{r_x} $, and $ C_\perp^\top C_\perp = I_{r_y} $. Additionally, $ [WW_\perp] $ and $ [CC_\perp] $ do not have linearly dependent columns. Furthermore, the covariance matrices for the latent variables $ \Sigma_t, \, \Sigma_u, \, \Sigma_{t_\perp}, \, \Sigma_{u_\perp} $ are diagonal. Finally, the diagonal elements of $ B $ are positive and the diagonal elements of $ \Sigma_t B $ are strictly decreasing.

We show that these conditions also guarantee the identifiability (up to sign) of the GLM-PO2PLS model.
\begin{theorem} \label{theorem1}
	Let $(x,y,z)$ follow the GLM-PO2PLS model where $z$ is normally distributed. Additionally, let the parameters satisfy the PO2PLS conditions as described above.
	It follows that the GLM-PO2PLS model parameters are identifiable up to a sign.
\end{theorem}
\begin{proof}
	Let $ f(x,y,z|\theta) = f(x,y,z|\tilde{\theta}) $ be identical joint distributions under two sets of parameters $ \theta $ and $ \tilde{\theta} $. Then we necessarily have $ f(x,y|\theta) = f(x,y|\tilde{\theta}) $. Since $ (x,y|\theta) $ follows a zero mean multivariate normal distribution, its distribution is uniquely defined by the covariance matrix $ \Sigma_{x,y|\theta} $. Thus $ \Sigma_{x,y|\theta} = \Sigma_{x,y|\tilde{\theta}} $ follows. It has been proven in~\cite{Bouhaddani2021} that if $ \Sigma_{x,y|\theta} = \Sigma_{x,y|\tilde{\theta}} $ holds, then the parameters involved (i.e., $ \{W,C,W_\perp, C_\perp, B, \Sigma_t, \Sigma_{t_\perp}, \allowbreak \Sigma_{u_\perp}, \sigma_e^2, \sigma_f^2, \Sigma_h\} $) are identified, up to sign.
	
	For a normally distributed $z$, the random vector $ (x,y,z) $ follows a zero mean multivariate normal distribution, and its distribution is uniquely defined by the covariance matrix $ \Sigma_{\theta} $ in~\eqref{For:sigtheta}. It follows from $ f(x,y,z|\theta) = f(x,y,z|\tilde{\theta}) $ that $ \Sigma_{\theta} = \Sigma_{\tilde{\theta}} $. Now let $ a \Sigma_t W^\top = \tilde{a} \tilde{\Sigma_t} \tilde{W}^\top $. Since $ \Sigma_t W^\top =  \tilde{\Sigma_t} \tilde{W}^\top $ and is of full rank, we have $ a =\tilde{a} $. Similarly, we have $ b = \tilde{b} $, and $ \sigma_g^2 = \tilde{\sigma_g^2} $ from $ b\Sigma_h C^\top = \tilde{b} \tilde{\Sigma_h} \tilde{C}^\top $ and $ a\Sigma_t a^\top + b\Sigma_h b^\top + \sigma_g^2 = \tilde{a}\tilde{\Sigma_t} \tilde{a}^\top + \tilde{b}\tilde{\Sigma_h} \tilde{b}^\top + \tilde{\sigma_g^2} $, respectively. This shows identifiability of all the parameters in $ \theta $. 
\end{proof}

\subsubsection{Maximum likelihood estimation}
Since the GLM-PO2PLS model is a latent variable model and the likelihood factorizes in terms which can be maximized separately, we propose an EM algorithm~\cite{Dempster1977} to obtain maximum likelihood estimates of the model parameters.  

Suppose we observe the $(x,y,z)$ for $N$ subjects. Since we assume a multivariate normal distribution of $ (x,y,z) \sim \mathcal{N}(0,\Sigma_\theta) $,
the log-likelihood for one subject is given by
\begin{equation*}
	\ell(\theta; x,y,z) = -\frac{1}{2}\{ (p+q+1) \log(2\pi) + \log|\Sigma_\theta| + (x,y,z) \Sigma_\theta^{-1} (x,y,z)^\top \}.
\end{equation*}

Denote the complete data vector by $ (x, y, z, t, u, t_\perp, u_\perp) $. For each current estimate $ \theta' $, the
EM algorithm considers the objective function
\begin{equation*}
	Q(\theta|\theta') = \mathbb{E}[\log  f(x,y,z,t,u,t_\perp, u_\perp | \theta) | x,y,z,\theta'].
\end{equation*}

\paragraph*{Expectation step}
The conditional expectation of the complete data log likelihood can be decomposed into different terms, 
\begin{equation}\label{For:EM_coml}
	\begin{aligned}
		\begin{split}
			Q(\theta|\theta') = & \E[\log f(x,y,z,t,u,t_\perp, u_\perp)] 
			= \E[\log  f(x,y,z | t,u,t_\perp, u_\perp)] + \E[\log  f(t,u,t_\perp, u_\perp)] \\
			= &  \undertext{\E[\log  f(z|t,u)]}{Q_{\{a,b,\sigma_g^2\}}}
			+ \undertext{\E[\log  f(x|t,t_\perp)]}{Q_{\{W,W_\perp,\sigma_e^2\}}} 
			+ \undertext{\E[\log  f(y|u,u_\perp)]}{Q_{\{C,C_\perp,\sigma_f^2\}}} \\	
			& +\undertext{\E[\log  f(u|t)]}{Q_{\{B,\Sigma_h\}}}	
			+\undertext{\E[\log  f(t)]}{Q_{\Sigma_t}}
			+\undertext{\E[\log  f(t_\perp)]}{Q_{\Sigma_{t_\perp}}}
			+\undertext{\E[\log  f(t_\perp)]}{Q_{\Sigma_{u_\perp}}}.
		\end{split}
	\end{aligned}
\end{equation}
In this equation, the conditioning on $x$, $y$, $z$ and $\theta'$ is dropped, to simplify notation. 
The individual conditional expectations depend on distinct sets of parameters, yielding separate optimization tasks. Compared to PO2PLS, the extra parameters in GLM-PO2PLS $ \{a,b,\sigma_g^2\} $ are included in the first term $ Q_{\{a,b,\sigma_g^2\}} $. Therefore, we focus on the optimization of $ Q_{\{a,b,\sigma_g^2\}} $ with respect to $ \{a,b,\sigma_g^2\} $. The rest of the terms are identical to the factorized densities in the original PO2PLS EM algorithm, we refer to the PO2PLS paper~\cite{Bouhaddani2021} for the expectation and maximization regarding these terms.

In the expectation step, $ Q_{\{a,b,\sigma_g^2\}} $ is calculated as
\begin{equation}\label{For:Qz}
	Q_{\{a,b,\sigma_g^2\}} = -\frac{1}{2}\bigg\{\log(2\pi \sigma_g^2) + \frac{1}{\sigma_g^2} \tr \mathbb{E}\left[ (z-ta^\top-(u-tB)b^\top)^\top(z-ta^\top-(u-tB)b^\top) \right]\bigg\}.
\end{equation}
Here, the first and second conditional moments of the vector $ (t,u) $ given $ x,y,z $ and $ \theta' $ are involved. Since $ (x,y,z,t,u,t_\perp, u_\perp) $ follows a multivariate normal distribution with zero mean and known covariance matrix, the conditional density $ f(t,u,t_\perp, u_\perp | x,y,z) $ can be calculated following Lemma 3 in~\cite{Bouhaddani2021}. The conditional moments involved in~\eqref{For:Qz} can then be obtained from the mean and the covariance matrix of $ (t,u,t_\perp, u_\perp | x,y,z) $ (see the Supplementary material for details). 

\paragraph*{Maximization step}
In the maximization (M) step, each conditional expectation in~\eqref{For:EM_coml} can be optimized separately. Here, we restrict to the description of the term involving the outcome, namely, maximize the $ Q_{\{a,b,\sigma_g^2\}} $ as given in equation~\eqref{For:Qz}. Note that the coefficient vector $ (a,b) $ can be separately optimized from the residual parameter $ \sigma_g^2 $, as in the standard linear regressions.
We first calculate the derivative with respect to $ (a,b) $ and set it to 0, yielding
\begin{equation*}
	\frac{\partial Q_{\{a,b,\sigma_g^2\}}}{\partial (a,b)} = 0 \Rightarrow (\hat{a},\hat{b}) = z^\top \E[(t,h)] \E[(t,h)^\top (t,h)]^{-1}.
\end{equation*}
where the conditional moments are calculated in the E step. The maximization with respect to the parameter $ \sigma_g^2 $ can then be performed similarly. Details are given in the supplementary material.

\subsubsection{Statistical inference}
The GLM-PO2PLS method allows for statistical inference on the relationship between the omic data and the outcome. 
This relationship is captured by the joint parts $t$ and $u$, and given by the equation $ \eta(\mathbb{E}[z]) = ta^\top + ub^\top $ in~\eqref{For:model}. Here, we propose two tests, one full test for the relationship between $ z $ and all the joint components together, and one component-wise test for the relationship between $ z $ and each pair of joint components.

For the full test, we consider the null hypothesis,
\begin{equation*}
	H_0 \, : \, a = b = \mathbf{0}  \qquad  \text{against}  \qquad H_1 \, : \, a \neq \mathbf{0} \; \text{or} \; b \neq \mathbf{0}.
\end{equation*} 
For each component-wise test, we consider the null hypothesis of no relationship between $ z $ and the $ k $-th pair of joint components,
\begin{equation*}
	H_0 \, : \, a_k = b_k = 0  \qquad  \text{against}  \qquad H_1 \, : \, a_k \neq 0 \; \text{or} \; b_k \neq 0.
\end{equation*} 
where $ a_k $ and $ b_k $ are the coefficients for $ t_k $ and $ h_k $, respectively.

Let $ \alpha=(a,b) $ and $ \alpha_k=(a_k,b_k) $. The full test statistic is given by
\begin{equation} \label{Testchi}
	T_{full} = \hat{\alpha} \Pi_{\hat{\alpha}}^{-1} \hat{\alpha}^\top,
\end{equation}
where $ \Pi_{\hat{\alpha}}^{-1} $ is the inverse of the covariance matrix of $ \hat{\alpha} $. And the pair-wise test statistic is given by:
\begin{equation} \label{Testchi2}
	T_{comp.wise} = \hat{\alpha_k} \Pi_{\hat{\alpha_k}}^{-1} \hat{\alpha_k}^\top.
\end{equation}

To calculate the (asymptotic) distribution of these test statistics, the asymptotic distribution of all parameters $ \theta $ needs to be derived. 

\paragraph*{Asymptotic distribution}
Under certain regularity conditions, consistency of the estimator $ \theta $ and its asymptotic distribution $ \mathcal{N}(\theta, \Pi_\theta) $ follows from Shapiro’s Proposition 4.2 (Shapiro 1986) applied to the GLM-PO2PLS model. 
\begin{theorem}
	Let $\hat{\theta}$ be the maximum likelihood estimator for $\theta$ from the GLM-PO2PLS model. When the sample size $N$ approaches infinity, the distribution of $\hat{\theta}$ converges to a normal distribution, i.e.
	\begin{equation*}
		N^{1/2}(\hat{\theta} - \theta) \longrightarrow \mathcal{N}(0,\Pi_\theta)
	\end{equation*}
\end{theorem}
Details and proofs are given in the supplement. 

In particular, $\hat{\alpha} = (\hat{a}, \hat{b})$ is asymptotically normally distributed. Therefore, the test statistics $T_{full}$ and $T_{comp.wise}$ follow a chi-square distribution with $2r$ resp. $2$ degrees of freedom. An estimate of $ \Pi_\theta $ is obtained from the inverse observed Fisher information matrix. Let $ \psi_i $ be an instance of observed data $ (x,y,z) $ and $ \zeta_i $ be the latent variables involved. In an EM algorithm, this matrix is given by~\cite{Louis1982}, 
\begin{equation*}
	\begin{aligned}
		\mathcal{I}(\hat{\theta}) = & \sum_{i=1}^{N} \E[B_i(\hat{\theta})|\psi_i  ]
		- \sum_{i=1}^{N} \sum_{j=1}^{N} \E[S_i(\hat{\theta}) S_j(\hat{\theta})^\top |\psi_i  ; \psi_j  ]
	\end{aligned}
\end{equation*}
where $ S_i(\hat{\theta}) = \nabla l(\hat{\theta};\psi_i, \zeta_i) $ and $ B_i(\hat{\theta}) = -\nabla^2 l(\hat{\theta};\psi_i, \zeta_i) $ are the gradient and negative of the second derivative of the log complete likelihood of instance $ i $, respectively, evaluated at $ \hat{\theta} $.

To obtain $ \Pi_{\hat{\alpha}} $, the submatrix of $ \mathcal{I}^{-1}(\hat{\theta}) $ corresponding to $ \hat{\alpha} $ (denote $ \mathcal{I}^{-1}(\hat{\alpha}) $) has to be calculated. However, inverting $ \mathcal{I}(\hat{\theta}) $ is computationally infeasible, even for moderate dimensions. Under additional assumptions that $ \hat{\alpha} $ and $ \hat{\theta}/\hat{\alpha} $ are asymptotically independent and $ \hat{\sigma}_g^2 $ is non-random, $ \mathcal{I}^{-1}(\hat{\alpha}) $ can be calculated, and be used to approximate $ \Pi_{\hat{\alpha}} $. The details are given in supplementary materials.

\subsection{The GLM-PO2PLS model with a binary outcome}
For a binary outcome, we use a Bernoulli distribution for $z$ and the logit link function $\eta(v) = \logit(v) = \log[v(1-v)^{-1}]$. The model is then given by
\begin{flalign*}
	\begin{aligned}
		x = tW^\top + t_\perp W_\perp^\top + e, \quad
		y & = uC^\top + u_\perp C_\perp^\top + f, \quad
		u = tB + h, \\
		\logit (p(z)) & = a_0 + ta^\top + hb^\top.
	\end{aligned}
\end{flalign*}
Here, $ p(z) = \text{Pr}(z=1|t,h) $ is the conditional probability of the random variable $ z $ being 1, given $ t $ and $ h $. Note that the probability $p(z)$ is logit-normally distributed, therefore the linear predictor $\logit(p(z))$ follows a normal distribution $ \mathcal{N}(a_0, a\Sigma_t a^\top + b\Sigma_h b^\top) $. The joint distribution $(x,y,\logit(p(z)))$ is multivariate normal with mean vector $ (0_{p+q}, a_0) $ and covariance matrix $\Sigma_\theta$ in~\eqref{For:sigtheta} excluding $\sigma_g^2$. The collection of parameters in the GLM-PO2PLS model with a binary outcome is $ \theta = \{W,C,W_\perp, C_\perp, a_0, a, b, B, \Sigma_t, \Sigma_{t_\perp}, \allowbreak \Sigma_{u_\perp}, \sigma_e^2, \sigma_f^2, \Sigma_h\} $. 

\paragraph*{Identifiability of GLM-PO2PLS with a binary outcome} Theorem~\ref{theorem1} also appears to hold for a binary $ z $ that follows a Bernoulli distribution, under the same conditions. The proof is similar. Specifically, let $ f(x,y,z|\theta) = f(x,y,z|\tilde{\theta}) $ be identical joint distributions under two sets of parameters $ \theta $ and $ \tilde{\theta} $. Then $ f(x,y|\theta) = f(x,y|\tilde{\theta}) $, thus $ \Sigma_{x,y|\theta} = \Sigma_{x,y|\tilde{\theta}} $ holds regardless of the distribution of $ z $. The conclusion follows that the parameters involved in PO2PLS model (i.e., $ \{W,C,W_\perp, C_\perp, B, \Sigma_t, \Sigma_{t_\perp}, \allowbreak \Sigma_{u_\perp}, \sigma_e^2, \sigma_f^2, \Sigma_h\} $) are identified up to sign.
Now consider $(x,y,\logit(p(z)))$ which is multivariate normally distributed with mean vector $ (0_{p+q}, a_0) $ and covariance matrix $\Sigma_\theta$ excluding $\sigma_g^2$ (denote $\Sigma_{\theta/g^2}$). Since the mapping $ f(x,y,z|\theta) \mapsto f(x,y, \logit(p(z))|\theta) $ is one-to-one, it follows that $ f(x,y,\logit(p(z))|\theta) = f(x,y,\logit(p(z))|\tilde{\theta}) $. Necessarily, the means and covariance matrices of two identical multivariate normal distributions are equivalent, thus $ (0_{p+q}, a_0) = (0_{p+q}, \tilde{a}_0) $ and $ \Sigma_{\theta/g^2} = \Sigma_{\tilde{\theta}/g^2} $. It is clear that $ a_0= \tilde{a}_0 $ from the equivalence of the mean vectors. The identifiablity of $ a $ and $ b $ can be shown from the equivalence of covariance matrices analogously as in the proof of Theorem~\ref{theorem1}. This shows the identifiability of all the parameters in GLM-PO2PLS with a binary outcome. 

\subsubsection{EM algorithm for a binary outcome}
For a Bernoulli distributed outcome, the log-likelihood of the observed data involves an integral of dimension $ 2r+r_x+r_y $. Let $ \nu = (t,u) $ and $ \xi = (t_\perp, u_\perp) $,
\begin{equation}\label{For:lbi}
	\ell(\theta; x,y,z) = \log  \int_{(\nu, \xi)} f(x,y,z|\nu, \xi, \theta) f(\nu, \xi|\theta) d(\nu, \xi).
\end{equation}
To estimate~\eqref{For:lbi}, numerical integration is needed. 
Note that given $ \nu $, the binary outcome $ z $ is independent of $x$, $y$ and $ \xi $, thus the conditional density $ f(x,y,z|\nu, \xi) $ in~\eqref{For:lbi} can be factorized as $ f(x,y,z|\nu, \xi) = p(z|\nu) f(x,y|\nu,\xi) $. The factorization enables to integrate out the specific random  vector $ \xi $, hence reducing the dimension of the integral to $ 2r $,
\begin{equation*}
	\begin{aligned}
		\ell(\theta; x,y,z) 
		& =  \log  \int_{(\nu, \xi)}  p(z|\nu) f(x,y|\nu,\xi) f(\nu,\xi) d(\nu, \xi) \\
		& = \log  \int_\nu p(z|\nu) \left[\int_\xi f(x,y|\nu,\xi) f(\xi |\nu) d\xi \right] f(\nu) d\nu\\
		& = \log  \int_\nu p(z|\nu) f(x,y|\nu) f(\nu) d\nu\\
		& = \log  \int_\nu p(z|\nu) f(x|\nu) f(y|\nu) f(\nu) d\nu.\\
	\end{aligned}
\end{equation*}
Here, the probability mass function $ p(z|\nu) $ is given by
\begin{equation} \label{conpz}
	p(z|\nu) = 
	\begin{dcases}
		\Big(1+\exp\{-(a_0+ta^\top+(u-tB)b^\top)\}\Big)^{-1} & z = 1,\\
		\Big(1+\exp\{a_0+ta^\top+(u-tB)b^\top\}\Big)^{-1} & z = 0.
	\end{dcases} 
\end{equation}
The probability density functions $ f(x|\nu) $, $ f(y|\nu) $, and $ f(\nu) $ follow from the following multivariate normal distributions, 
\begin{equation*}
	x|\nu \sim \mathcal{N}(tW^\top, \Sigma_{x|t}), \qquad y|\nu \sim \mathcal{N}(uC^\top, \Sigma_{y|u}), \qquad  \nu \sim \mathcal{N}(0, \Sigma_\nu)
\end{equation*}
where the covariance matrices involved are:
\begin{equation*}
	\Sigma_{x|t}= W_\perp \Sigma_{t_\perp} W_\perp^\top + \sigma_e^2 I_p,
	\qquad
	\Sigma_{y|u}= C_\perp \Sigma_{u_\perp} C_\perp^\top + \sigma_f^2 I_q,
	\qquad
	\Sigma_{\nu}=
	\begin{bmatrix}
		\Sigma_t	&	\Sigma_t B \\
		B\Sigma_t	&	\Sigma_u
	\end{bmatrix}.
\end{equation*}

Denote the partial complete data vector by $ (x, y, z, \nu) $. For each current estimate $ \theta' $, the
EM algorithm for a binary outcome considers the objective function
\begin{equation}\label{For:Qbz}
	Q(\theta|\theta') = \mathbb{E}[\log  f(x,y,z,\nu| \theta)| x,y,z,\theta'].
\end{equation}

\paragraph*{Expectation step based on numerical integration}
Analogously to~\eqref{For:EM_coml}, the conditional expectation in~\eqref{For:Qbz} can be decomposed to factors that depend on distinct sets of parameters,
\begin{equation}\label{For:EM_parcoml}
	\begin{aligned}
		\begin{split}
			Q(\theta|\theta') & = \E[\log f(x,y,z,\nu)] 
			= \E[\log  f(x,y,z | \nu)] + \E[\log  f(\nu)] \\
			& = \undertext{\E[\log  p(z|\nu)]}{Q_{\{a_0,a,b\}}}
			+ \undertext{\E[\log  f(x|t)]}{Q_{\{W,W_\perp,\sigma_e^2,\Sigma_{t_\perp}\}}} 
			+ \undertext{\E[\log  f(y|u)]}{Q_{\{C,C_\perp,\sigma_f^2,\Sigma_{u_\perp}\}}}	
			+\undertext{\E[\log  f(u|t)]}{Q_{\{B,\Sigma_h\}}}	
			+\undertext{\E[\log  f(t)]}{Q_{\Sigma_t}}. \\
		\end{split}
	\end{aligned}
\end{equation}
Here, the first conditional expectation $ Q_{\{a_0,a,b\}} $ has no closed form,
\begin{equation*}
	\begin{aligned}
		Q_{\{a_0,a,b\}} 
		= \int [\log  p(z|\nu)] f(\nu|x,y,z,\theta') d\nu
		= \frac{1}{f(x,y,z)} \int [\log  p(z|\nu)] p(z|\nu) f(x,y|\nu) f(\nu) d\nu.
	\end{aligned}
\end{equation*}
To obtain an approximation of the multivariate integral, Gauss–Hermite quadrature can be used. For an integral of form $ \int \varphi(\nu) p(z|\nu) f(x,y|\nu) f(\nu) d\nu $, where $ \varphi $ is any function, we approximate it with
\begin{equation} \label{GH}
	\begin{aligned}
		\int \varphi(\nu) p(z|\nu) f(x,y|\nu) f(\nu) d\nu 
		\approx \sum_{m_1=1}^M \ldots \sum_{m_{2r}=1}^M \varphi(\nu = \nu_m) p(z|\nu = \nu_m) f(x,y|\nu = \nu_m) w_{m_1}\ldots w_{m_{2r}}
	\end{aligned}
\end{equation}
with nodes vector $ \nu_m = (\nu_{m_1}, \ldots \nu_{m_{r}}) = \sqrt{2}(\Sigma_{\nu}^{1/2})^\top \nu_m^* $ and weights vector $ w_m = (w_{m_1}, \allowbreak \ldots w_{m_{r}}) = w_m^*/\sqrt{\pi} $. Here, $ M $ is the number of sampling nodes, $ \Sigma_{\nu}^{1/2} $ is the Cholesky decomposition of $ \Sigma_{\nu} $, and $ \nu_m^* $ and $ w_m^* $ are nodes and weights of a $ M $-point standard Gauss–Hermite quadrature rule, which can be found on Page 924 in~\cite{Abramowitz1972}. The transformation from the standard quadrature nodes $ \nu_m^* $ to $ \nu_m $ is to make the sampling range of the integrand in~\eqref{GH} more suitable based on the distribution of $ \nu $~\cite{Liu1994}.

The other terms in~\eqref{For:EM_parcoml} have explicit expressions in terms of the first and second conditional moments of the vector $ \nu $ given $ x,y,z $ and $ \theta' $ (see for details in the Supplementary materials). Note that the conditional moments of $ \nu $ are in forms of integrals as follows
\begin{equation*}
	\begin{aligned}
		\E[\nu|x,y,z,\theta'] & =\int \nu f(\nu|x,y,z)d\nu = \frac{1}{f(x,y,z)} \int \nu p(z|\nu)f(x,y|\nu)f(\nu) d\nu, \\
		\E[\nu^\top \nu|x,y,z,\theta'] & =\int \nu^\top \nu f(\nu|x,y,z)d\nu = \frac{1}{f(x,y,z)} \int \nu^\top \nu p(z|\nu)f(x,y|\nu)f(\nu)d\nu,
	\end{aligned}
\end{equation*}
which can be numerically calculated with~\eqref{GH}.

\paragraph*{Maximization step based on gradient descent} 
Maximizing $ Q_{\{a_0,a,b\}} $ requires iterations as its derivative with respect to $ \beta = (a_0, a, b) $ has no analytical solutions. To find an update of $ \beta $ in each EM iteration,  we propose a one-step gradient descent strategy. The gradient of $ Q_\beta $ is given by
\begin{equation*}
	\begin{aligned}
		\nabla Q_\beta & = 	\Big[\frac{\partial Q_\beta}{\partial \beta}\Big]^\top
		= \Big[ \frac{1}{f(x,y,z)} * \frac{\partial}{\partial \beta}  \int [\log  p(z|\nu)] p(z|\nu) f(x,y|\nu) f(\nu) d\nu  \Big]^\top \\
		& = \Big[ \frac{1}{f(x,y,z)} \int \frac{\partial \log  p(z|\nu)}{\partial \beta} p(z|\nu) f(x,y|\nu) f(\nu) d\nu \Big]^\top
	\end{aligned}
\end{equation*}

To guarantee the increase of $ Q_\beta $ in each EM iteration, we search for a step size along the direction of the gradient using the backtracking rule (also known as the Armijo rule)~\cite{Armijo1966}. It is performed by starting with an initial step size of $ s = 1 $ for movement along the gradient, and iteratively shrinking the step size ($ s \leftarrow 0.8* s $) until an increase of $ Q_\beta $ exceeds the expected increase based on the local gradient. More precisely, we keep shrinking the step size until the following ascent condition is met:
\begin{equation*} 
	\begin{aligned}
		Q_{(\beta + s\nabla Q_{\beta})} \geq Q_{\beta} + 0.5 * s \nabla Q_{\beta} \nabla Q_{\beta}^\top.
	\end{aligned}
\end{equation*}

The maximization of the other conditional expectation terms in~\eqref{For:EM_parcoml} can be found in the supplementary materials.

	\section{Simulation}
We conduct a simulation study to evaluate the performance of GLM-PO2PLS.
Both continuous outcome $ z_c $ and binary outcome $ z_b $ are investigated. The datasets are simulated following the GLM-PO2PLS model in~\eqref{For:model}, with the equations for the continuous and binary outcomes being $ z_c = ta^\top + hb^\top + g $, and $ z_b \sim \text{Bernoulli}((1+\exp\{-(a_0+ta^\top+hb^\top)\})^{-1}) $.

\subsection{Simulation settings}
We consider combinations of small and large sample sizes $ (N = 100, 1000) $ with low and high dimensionalities $ (p = 100, 2000; q = 10, 25) $.  The latent variables $ t, \, t_\perp, \, u_\perp $ are simulated from standard normal distribution, and $ u=tB+h $ following equation~\eqref{For:model}. Here, $ B $ is the identity matrix and the joint residual $ h $ in $ u $ that is independent of $ t $ determines the level of heterogeneity in the joint parts. To investigate the impact of heterogeneity levels, we vary the variance of $ h $ to account for 40\% and 80\% of the total variance in $ u $. The residual terms $ e, \, f $ are generated from zero-mean normal distributions. In the low noise level scenario, we set the noise proportion in $ x $ and $ y $ to both 40\%. In the high noise level scenario, we investigate the performance of GLM-PO2PLS when integrating a very noisy large dataset and a less noisy small dataset, by increasing the noise in $ x $ to 95\% and decreasing the noise in $ y $ to 5\%. The noise term $ g $ for the continuous outcome is generated from a zero-mean normal distribution, accounting for 20\% of variation in $ z_c $.
All the loading matrices are generated from standard normal distribution and then semi-orthogonalized. The coefficients $ a $ and $ b $ are set to 2 and 1, respectively. 
The number of joint and specific components is set to 1 for simplicity of the model and computational efficiency. For each setting, 500 replications are generated. The settings are summarized in Table~\ref{tbl:simu_para}. 

\begin{table}[htb]
	\centering
	\tbl{Summary of simulation settings}
	{\begin{tabular}{*3c}
		\toprule
		Notations &  Description & Setting/Distribution \\
		\midrule
		\multirow{2}{*}{$ N $} & \multirow{2}{*}{Sample size}	&  Small: 100 \\ 
		&	&Large: 1000 \\ \hline
		\multirow{2}{*}{$ p ; q $}	&	\multirow{2}{*}{ Dimension of $ x,y $}	& Low: 100,10 \\
		&	& High: 2000,25 \\ \hline
		\multirow{3}{*}{$ h $}	&	Heterogeneity between  	&Normal \\
		&joint latent variables	&Moderate: 40\% of variance in $ u $ \\
		&$ t $ and $ u $	&High: 80\% of variance in $ u $ \\ \hline
		\multirow{3}{*}{$ e,f $}	&	\multirow{3}{*}{Noise in $ x,y$}	&Normal \\
		&	&Low: 40\%, 40\%\\
		&	&High: 95\%, 5\%\\
		\bottomrule
	\end{tabular}}
	\label{tbl:simu_para}
\end{table}

The metrics used to assess the performance are listed in~Table~\ref{tbl:simu_metric}. We first study the estimation accuracy of the coefficients $ a $ and $ b $. The errors $ (\hat{a}-a) $ and $ (\hat{b}-b) $ are standardized by $ a $ and $ b $ to exclude the influence of the parameter scale. 
The performance of outcome prediction is assessed by root mean square error of prediction (RMSEP), defined as $ (\E[(\hat{z_c}-z_c)^2])^\frac{1}{2} $ for continuous outcome $ z_c $, and $ (\E[(\logit(p(z_b)) - \logit(\hat{p}(z_b)))^2])^\frac{1}{2} $ for binary outcome $ z_b $. We compare the performance of GLM-PO2PLS with ridge regression fitted separately on $ x $ (denote ridge-x) and on $ y $ (denote ridge-y). The shrinkage hyper-parameter in ridge regressions is searched using a 10-fold cross-validation for each fit. The prediction performance is evaluated on an independent test dataset of size 1000. The accuracy of loading estimation is measured by the inner product between the estimated and the true loading vectors. The performance of feature selection is measured by true positives rate (TPR) calculated as the proportion of true top 25\% features among the estimated top 25\% in $ x $ (i.e., the top 25\% of features in $ x $ with the largest absolute loading values in GLM-PO2PLS, or with the largest absolute regression coefficients in ridge regression).
\begin{table}[htb]
	\centering
	\tbl{Metric}
	{\begin{tabular}{*4c}
		\toprule
		Category	&	Metric	& Calculation	&	Competing methods \\
		\midrule
		Coefficient estimation	&	Scaled error	& $ (\hat{a}-a)/a $, $ (\hat{b}-b)/b $	& \\ \hline
		\multirow{2}{*}{Outcome prediction}	& 	\multirow{2}{*}{RMSEP}	& $ (\E[(\hat{z_c}-z_c)^2])^\frac{1}{2} $,	&	\multirow{2}{*}{ridge-x, ridge-y} \\
		&	&	$ (\E[(\logit(p(z_b)) - \logit(\hat{p}(z_b)))^2])^\frac{1}{2} $	& \\ \hline
		Loading estimation	&	Inner product	& $  W^\top\hat{W} $, $ W_\perp^\top\hat{W_\perp}$, $C^\top\hat{C}$,  $C_\perp^\top\hat{C_\perp} $	& \\ \hline
		Feature selection	&TPR of top 25\%	&	TP/(TP+FN)	& ridge-x \\
		\bottomrule
	\end{tabular}}
	\label{tbl:simu_metric}
\end{table}

\subsection{Results of simulation study}
\begin{figure*}[!h]
	\centering
	\subfloat[Continuous outcome $ z_c $.\label{fig:coef_su}]{\resizebox*{10cm}{!}{\includegraphics[width=\textwidth]{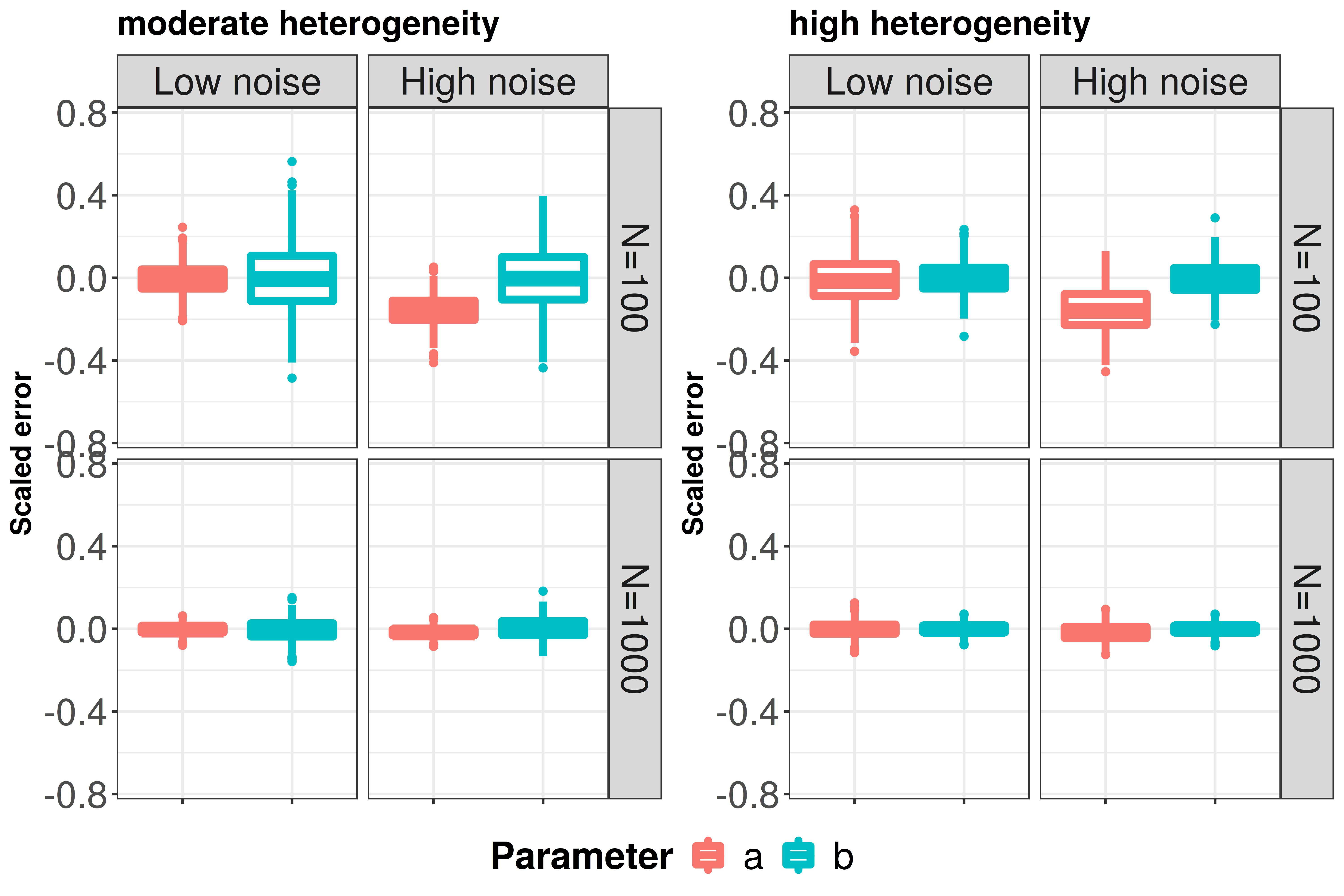}}}\\
	\subfloat[Binary outcome $ z_b $.\label{fig:coef_bi}]{\resizebox*{10cm}{!}{\includegraphics[width=\textwidth]{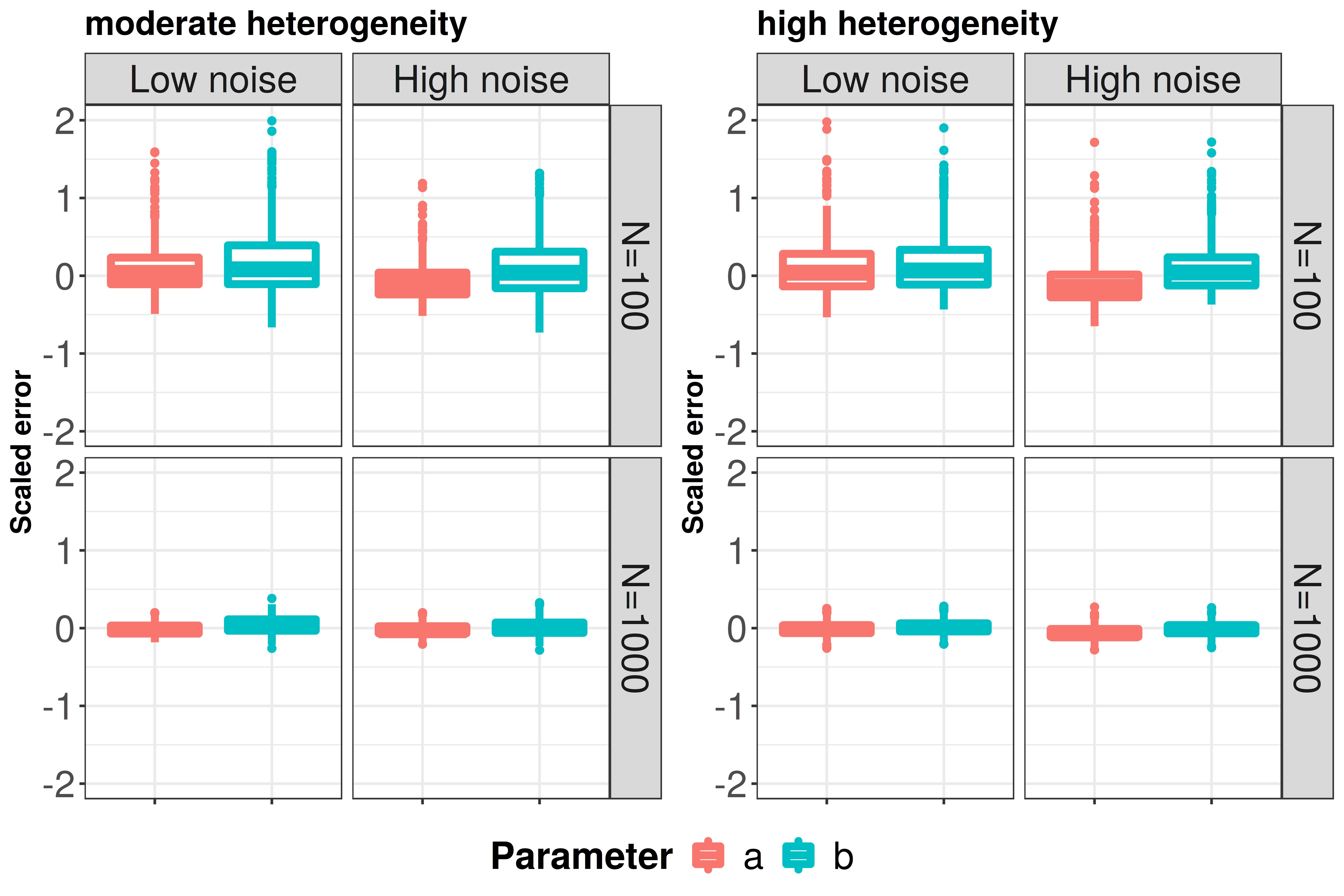}}}\\
	\caption{\textbf{Performance of coefficient estimation for continuous (a) and binary (b) outcome.} The y-axis shows the scaled estimation error as defined in Table~\ref{tbl:simu_metric}. In the moderate and high heterogeneity settings, $ h $ account for 40\% and 80\% of total variance in $ u=tB+h $, respectively. Boxes show the results of 500 repetitions.}
	\label{fig:coef}
\end{figure*}

In Fig~\ref{fig:coef}, results of the coefficient estimation in high-dimensional settings are depicted.
Fig~\ref{fig:coef_su} shows that for the continuous outcome, overall, the scaled errors of both $ \hat{a} $ and $ \hat{b} $  were small. When the sample size was small and the noise was high, the scaled error $ (\hat{a}-a)/a $ was mostly negative, suggesting that $a$ was underestimated.
For a large sample size, the estimators appeared to be unbiased. 
When the heterogeneity between the joint components was increased (from the left panel to the right), the joint residual $ h $ had larger variance relative to $ t $ and explained a larger proportion of $ z $. Consequently, the estimation of the coefficient $ b $ (for $ h $) became more stable, while the estimation of $ a $ (for $ t $) became less stable.
The results for a binary outcome are shown in Fig~\ref{fig:coef_bi}. Under a small sample size, the parameter estimation was less stable than the continuous case (note that the scale of y-axis in subplot (a) and (b) are different). The long upper whiskers suggested that the coefficients were overestimated in a some simulation runs. For a large sample size, the scaled errors for all coefficients were close to 0 and stable. 
Overall, the results for low dimensions were similar, except that the estimation of $ b $ was less stable in low dimensions compared to that in high dimensions. Details are given in the supplementary material.

\begin{figure*}[!h]
	\centering
	\subfloat[Continuous outcome $ z_c $.\label{fig:pre_su}]{\resizebox*{10cm}{!}{\includegraphics[width=\textwidth]{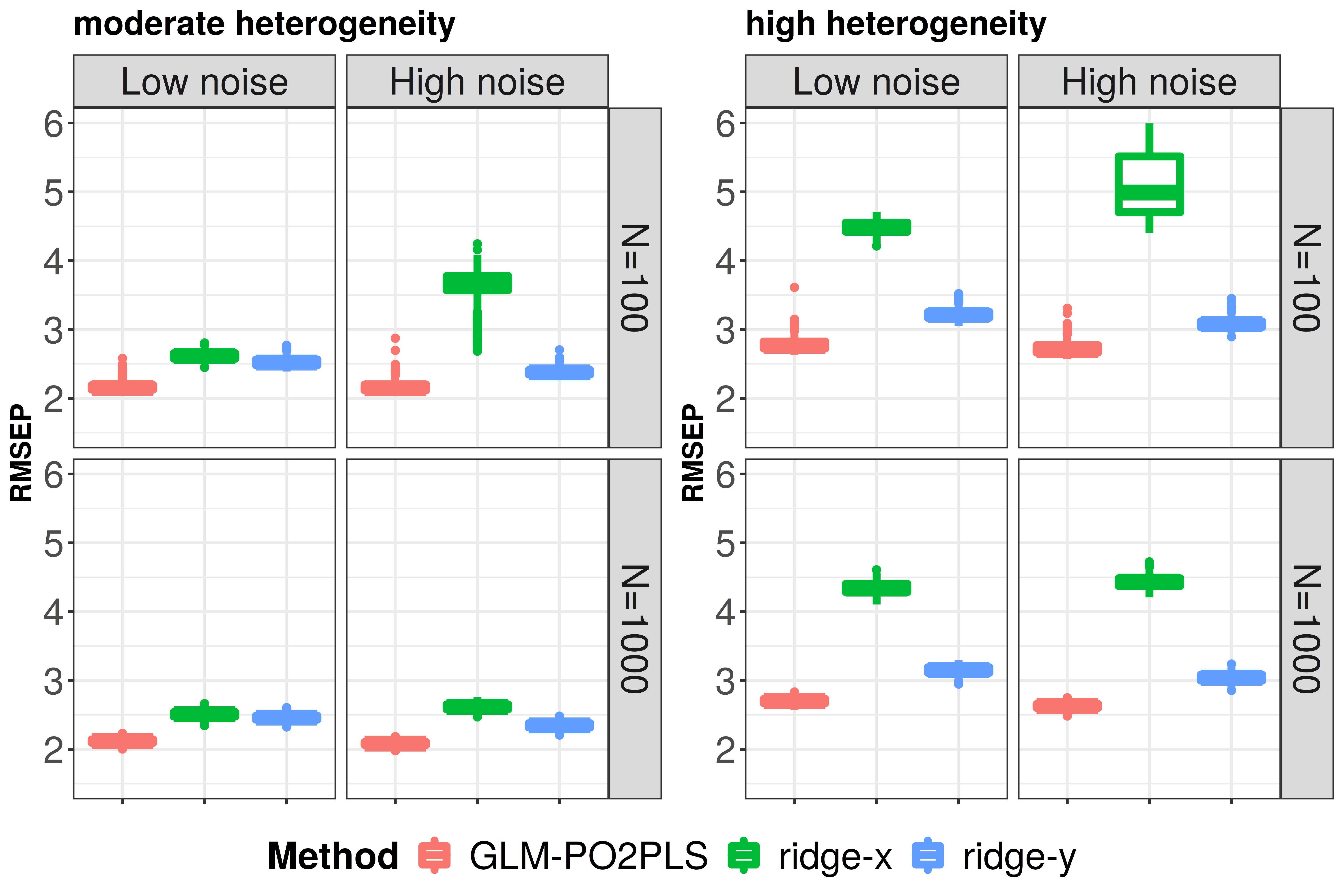}}}\\
	\subfloat[Binary outcome $ z_b $\label{fig:pre_bi}]{\resizebox*{10cm}{!}{\includegraphics[width=\textwidth]{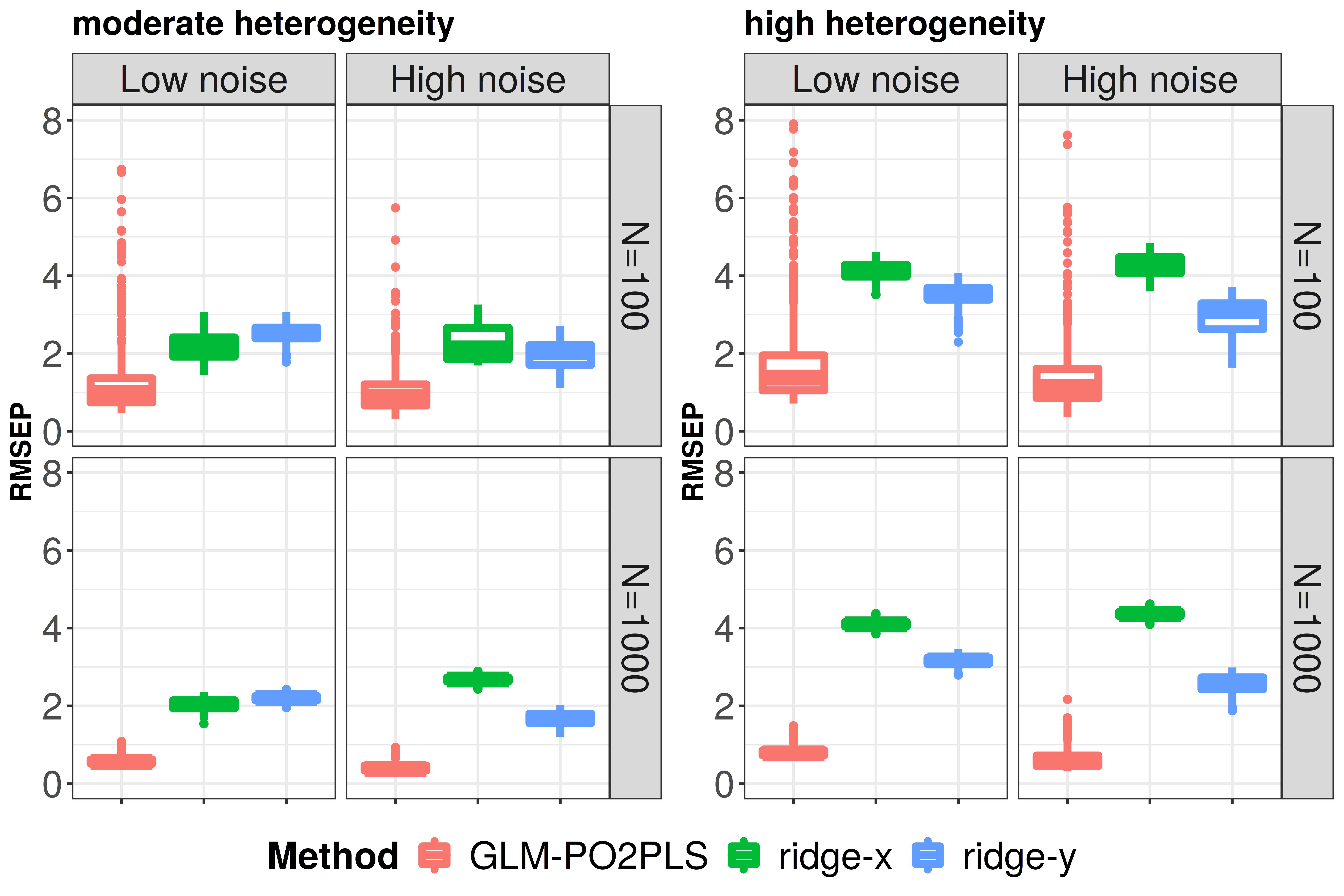}}}
	\caption{\textbf{Performance of outcome prediction for continuous (a) and binary (b) outcome.} y-axis shows the RMSEP as defined in Table~\ref{tbl:simu_metric}. Boxes show the results of 500 repetitions.}
	\label{fig:pre}
\end{figure*}

Fig~\ref{fig:pre} shows the results regarding outcome prediction in high-dimensional settings. For the continuous outcome, GLM-PO2PLS outperformed both ridge-x and ridge-y as shown in Fig~\ref{fig:pre_su}. The small boxes suggest that the prediction was very similar in each repetition, hence stable. Ridge-y performed similarly as GLM-PO2PLS, while ridge-x under-performed. When the noise in $x$ was increased, the performance of ridge-x deteriorated, especially when the sample size was small. The larger noise proportion in $ x $ barely affected the performance of GLM-PO2PLS. 
Increasing the heterogeneity made the RMSEP of ridge-x higher, as $ x $ explained less variation in $ z $, while the performance of GLM-PO2PLS was less affected.
For the binary outcome $ z_b $ , GLM-PO2PLS still outperformed ridge regression as shown in in Fig~\ref{fig:pre_bi}. 
When the sample size increased, the prediction of GLM-PO2PLS was less skewed and more stable. 
The conclusions also hold in low dimensions, details are given in the supplementary material.

Lastly, we briefly present the key results for loading estimation and feature selection. Overall, the loading estimates were accurate for both continuous and binary outcomes, with most inner products between the estimated and the true loadings approaching the optimum. 
When the sample size was small and the noise level was high, the accuracy of loading estimation for $ x $ dropped. This was the same setting in which $ \hat{a} $ was biased as is shown in Fig~\ref{fig:coef_su}. Regarding feature selection, the lowest median TPR of GLM-PO2PLS was 0.62 in the scenario with a small sample size, large noise proportion, and high heterogeneity. In the other scenarios, the median TPR was above 0.85. The details are given in the supplementary material.

\section{Application to Down syndrome study}
We apply the GLM-PO2PLS model to the Down syndrome dataset, aiming to investigate whether the relationship between methylation and glycomics is associated to DS, and select the relevant molecules involved in the relationship. Since Down syndrome is often considered as a model for aging~\cite{Horvath2015}, and both methylation and glycomics are associated with biological age~\cite{Horvath2013,Kristic2014}, we expect the DS patients to be more similar to their mothers than siblings.

\subsection{Data description}
The Down syndrome study includes 29 families. Each family consists of one Down syndrome patient (DSP), one non-affected sibling (DSS), and their mother (DSM). The family-based design is used to control for genetic and environmental influences.
Two DSS are missing. Thus, the total sample size $ N $ is equal to 85 . The ages of the DSPs range from 10 to 43, with a median of 24 years. The ages of the siblings are roughly matched with the DS patients, ranging from 14 to 52 years. The mothers have ages between 41 and 83, with a median of 57 years. 

For each individual, the whole blood methylation was measured using Infinium HumanMethylation450 BeadChip (Infinium 450k). After quality control following steps described in~\cite{Bacalini2015}, 450981 CpG sites were retained. Beta value was derived at each CpG site as the ratio of intensities between methylated and unmethylated alleles. White blood cell counts were estimated from the beta values and corrected for using R package `Meffil'~\cite{Min2018}. Age and sex were corrected for using multiple regression. The glycomic dataset consists of 10 plasma N-glycans measured using DNA sequencer-assisted fluorophore-assisted carbohydrate electrophoresis (DSA-FACE)~\cite{Borelli2015}. These glycans were logTA normalized~\cite{Uh2020} and corrected for age and sex.

We will fit a GLM-PO2PLS continuous model and a GLM-PO2PLS binary model to these data.  We set methylation as $ x $, glycomics as $ y $, and the DS status as $ z $. The direction from methylation to glycomics ($ x $ to $ y $) was chosen based on previous research~\cite{Wahl2018} that suggested the presence of an indirect influence of methylation on glycosylation.

\subsection{Results of DS data analysis}
For the GLM-PO2PLS continuous model, we used 3 joint and 1 specific component for each omic dataset based on the scree plots of the eigenvalues of $ x^\top y $, $ x^\top x $ and $ y^\top y $.

\begin{figure*}[!h]
	\centering
	\includegraphics[width=0.38\textwidth]{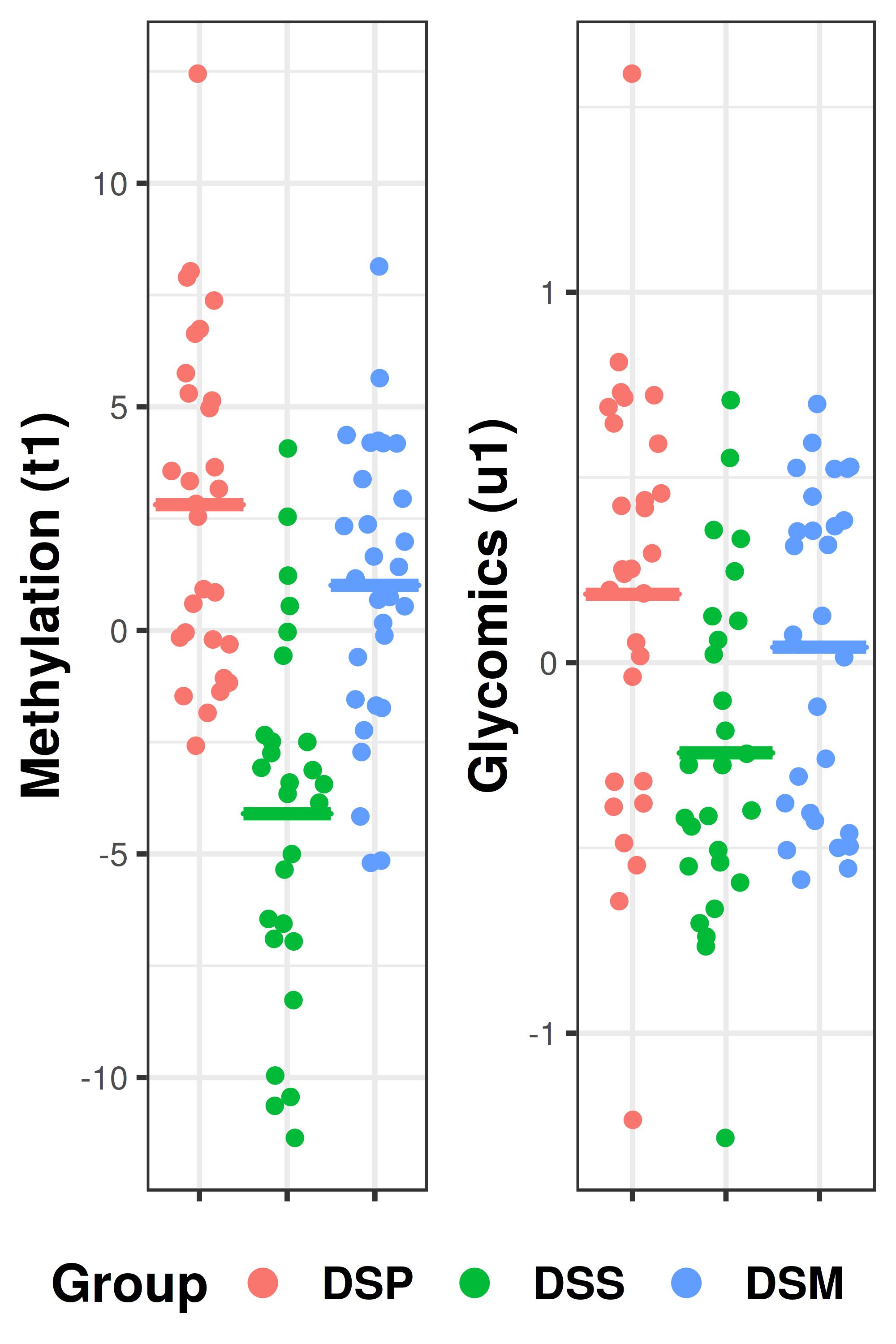} \qquad \qquad
	\includegraphics[width=0.38\textwidth]{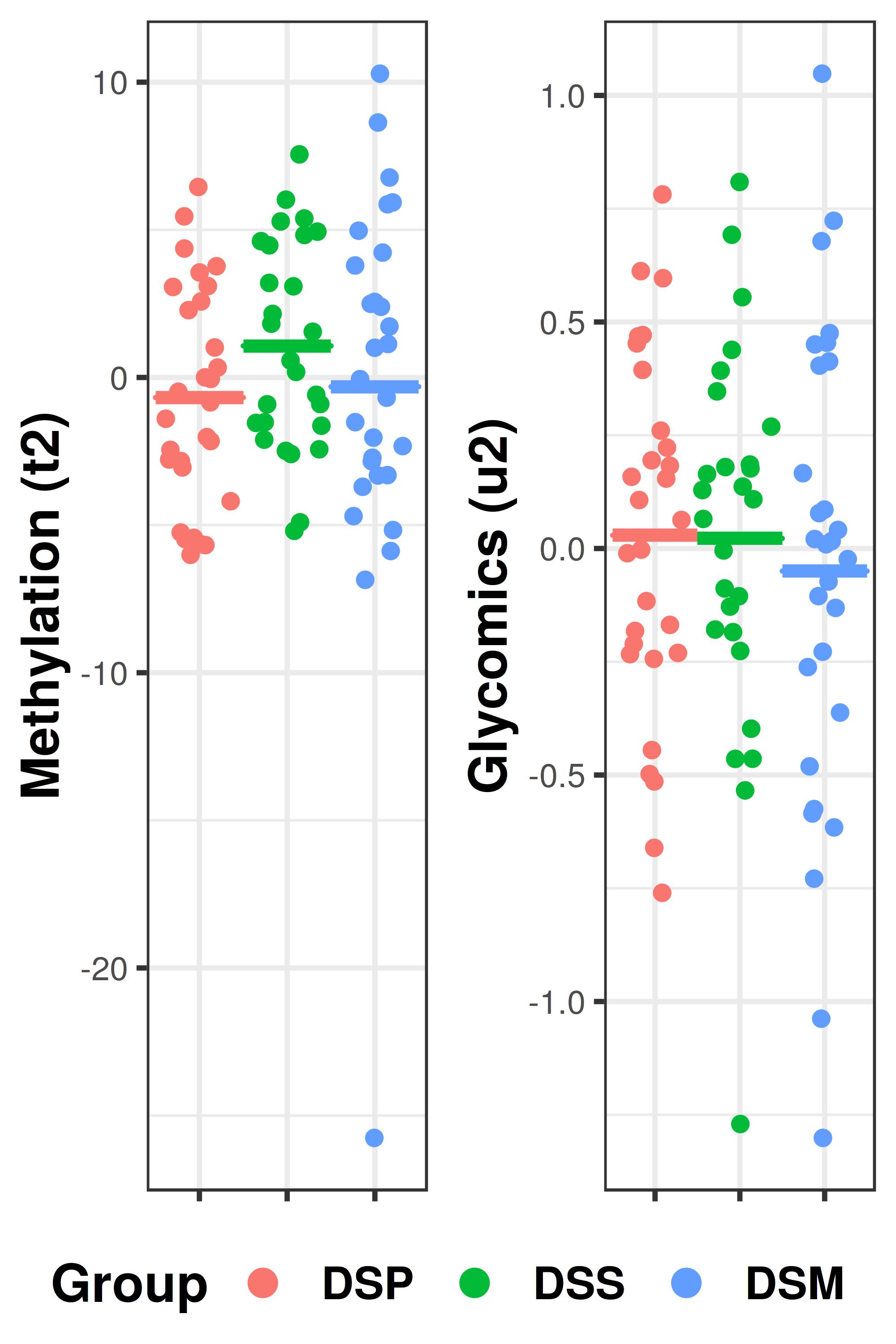}
	\caption{\textbf{Joint scores of the first (left) and second (right) pair of joint components.} On the y-axis are the scores of each individual colored by different groups. The mean score of each group is shown as a horizontal line.}
	\label{fig:js}
\end{figure*}

We first present the results regarding the relationship between methylation and glycomics, which is represented by the first three equations of the GLM-PO2PLS in~\eqref{For:model}.  The $ p $-value for each pair of methylation and glycomics joint components was 0.0007, 0.03, and 0.20, respectively. Using a threshold of 0.05 for statistical significance, the first ($ t_1 $ for methylation and $ u_1 $ for glycomics) and second pair ($ t_2 $ and $ u_2 $) of joint components were significantly associated. 
Fig~\ref{fig:js} shows the scores of the first two pairs of joint components. For both $ t_1 $ and $ u_1 $, the DSPs were closer to the DSMs, than the DSS group, which was in line with our expectation. No noticeable patterns were observed in the second pair of joint components.

Table~\ref{Table:tests} shows the results regarding the relationship between the DS status and the omics. The significant 
test statistic $ T_{full} $ suggests that DS was associated with the two omics. Component-wise, only the first pair was significant, with a $p$-value of $ 6.32 \times10^{-5} $.
\begin{table*}[htb]
	\centering
	\tbl{Results of testing for no relationship between DS and joint components}
{	\begin{tabular}{cc|ccc} 
		\toprule
		& $ T_{full} $ in~\eqref{Testchi}	&  \multicolumn{3}{c}{$ T_{comp.wise} $ in~\eqref{Testchi2}}\\
		\hline
		$ H_0 $ & $ a=b=0 $	& $ a_1=b_1=0 $ & $ a_2=b_2=0 $ & $ a_3=b_3=0 $ \\
		\hline
		$ p $-value &$ 6.32 \times10^{-5} $ & $ 1.35 \times10^{-5} $ & 0.15 & 0.20
	\end{tabular}}
	\label{Table:tests}
\end{table*}

Since $ t_1 $ and $ u_1 $ were significantly associated with DS, we investigated the CpG sites and glycans in the first component pair. In the first methylation joint component, the 1000 CpG sites with the largest loading values were mapped to their respective target genes, yielding 493 genes. Next, gene ontology (GO) enrichment analysis~\cite{Ashburner2000} was performed on this gene set using the GSEA software~\cite{Mootha2003,Subramanian2005}. The top three significant GO terms were listed in Table~\ref{Table:GO}. Among these terms, the cell-cell signaling is a biological function of plasma glycans~\cite{Varki2017,Krautter2021}. The cellular component of neuron projection and biological process of neurogenesis were shown to relate to DS~\cite{Haas2013,Huo2018,Stagni2018,Sobol2019}. 	
We further searched the mapped geneset in the DisGeNET database~\cite{Pinero2020} for human diseases. The significant diseases found were chronic myeloid leukemia (q-value 0.0004), common acute lymphoblastic leukemia (q-value 0.045), and glioblastoma multiforme (q-value 0.045). Research has shown that children with Down syndrome have an increased risk for developing acute lymphoblastic leukemia~\cite{Pinero2020}. For chronic myeloid leukemia and glioblastoma multiforme, we did not find evidence linking them with DS. 
We then checked the 7 genes with the highest GDA score regarding Down syndrome (i.e., with the most evidence of association with DS) in the DisGeNET database, and found the gene RCAN1 which relates to epigenetics was among our top genes mapped from methylation. It has been revealed that RCAN1 plays a critical upstream role in epigenetic regulation of adult neurogenesis~\cite{Choi2019}, hence important in the pathogenesis of Down syndrome~\cite{Yun2021}. 

For the first glycomics joint component, the glycan H3N4F1 had the largest absolute loading value. According to the result of a previous study~\cite{Borelli2015} on plasma glycans and DS, H3N4F1 was the top discriminators of DS subjects and siblings.

Next we fitted a GLM-PO2PLS binary model with 1 joint and 1 specific component for each omic dataset. We chose for 1 joint component based on the test results in the continuous model shown in Table~\ref{Table:tests}.
The relationship between the two omics was significant with a $ p $-value of 0.022.
The top 1000 CpG sites were identified and mapped to genes. The most significant GO terms of the geneset are shown in Table~\ref{Table:GO}. The top two terms were related to membrane organelle, more specifically, Golgi apparatus, which is required for accurate glycosylation~\cite{Zhang2016}. Terms related to DS, such as neurogenesis (q-value 2.33e-6), neuron differentiation (1.11e-5), and synapse (1.33e-5) were also significant. Regarding glycomics, the glycan with the largest absolute loading value was H3N4F1, which was also  identified in the GLM-PO2PLS continuous model.
\begin{table*}[!h]
	\centering
	\tbl{Top 3 GO terms of the mapped genesets in GLM-PO2PLS continuous and binary models}
	{\begin{tabular}{ |p{7cm}||p{2cm}|p{2cm}| } 
		\toprule
		\textbf{Gene Set Name (continuous model)} & $ p $-value & FDR q-value \\
		\hline
		GOBP CELL CELL SIGNALING  &  1.61e-13 &	2e-9	\\
		GOCC NEURON PROJECTION & 2.96e-13 &	2e-9 \\
		GOBP NEUROGENESIS & 4.08e-13 &	2e-9 \\
		\hline		
		
		\toprule
		\textbf{Gene Set Name (binary model)} & $ p $-value & FDR q-value \\
		\hline
		GOCC ORGANELLE SUBCOMPARTMENT  &  4.22e-12 &	4.3e-8	\\
		GOCC GOLGI APPARATUS & 8.27e-11 &	4.1e-7 \\
		GOCC VESICLE MEMBRANE & 1.21e-10 &	4.1e-7 \\
		\hline					
	\end{tabular}}
	\tabnote{The $ p $-value of each annotation was derived by random sampling of the whole genome; the FDR q-value provides the false discovery rate (FDR) analog of the p-value after correcting for multiple hypothesis testing~\cite{Benjamini1995,Storey2002}.}
	\label{Table:GO}
\end{table*}

Although for the continuous and binary models the sets of top CpG sites appeared to be relevant to glycosylation and DS, there was little overlap between the top  CpG sites. This might be explained by the different number of joint components specified for the two models. Therefore, we performed an additional  analysis using a filtered dataset, which was obtained by subtracting the second and third joint components, and the specific components of the GLM-PO2PLS continuous model from the omic data. The estimated parameters of the fitted GLM-PO2PLS binary model  were very similar to the ones in the continuous model. More specifically, the inner product of the joint loading vectors from the two models reached 0.99.  The correlation between the corresponding joint components of the two models was also high (at 0.99). Note that an inner product of 1 means the loading vectors are the same.
Regarding the interpretation of top CpG sites and glycans, we refer to the results in the continuous model, as the loadings were very similar, so as the top features identified.

\section{Discussion}	
Motivated by the studies on the	relationship among Down syndrome, methylation and glycomics, we developed a new statistical model GLM-PO2PLS, which simultaneously models the relationships among an outcome variable and two heterogeneous omic datasets.  We studied in detail the models for normally and Benoulli distributed outcome variables. The identifiability of the model was established and EM algorithms were developed. For testing, we proposed two chi-square test statistics $ T_{full} $ and $ T_{comp.wise} $ and derived their asymptotic distributions. 

Via a simulation study, we have shown that the model parameters were well estimated in various scenarios, and the outcome prediction performance of GLM-PO2PLS was robust against high noise and heterogeneity between omics. 
GLM-PO2PLS predicted the outcome better than ridge regressions, because it  considers all the information in the data jointly,  while ridge used each dataset separately. Another advantage of GLM-PO2PLS over ridge regression is that it can provide insights into the relationship between two omic datasets, on top of their relationship with the outcome.

The methylation and glycomics dataset were also analyzed by Bacalini et al.~\cite{Bacalini2015} and Borelli et al.~\cite{Borelli2015} using single point approaches for association with Down syndrome, respectively. Since for both omics datasets association with aging have been found and Down syndrome is an aging model, it makes sense to analyze them jointly in this study.
Concerning methylation, Bacalini et al. identified four categories of genes. Most of the genes in these categories were also in our obtained geneset:
haematopoiesis (RUNX1, DLL1, EBF4, PRDM16), morphogenesis and development (HOXA2, HOXA4, HHIP, NCAM1), neuronal development (NAV1, EBF4, PRDM8, NCAM1), and regulation of chromatin structure (PRDM8, KDM2B). 
In total four genes mentioned in~\cite{Bacalini2015} were not in our gene list, namely, HOXA5, TET1, GABBR1, and HOXA6. It appeared that three out of these four genes rank just below our cut-off point of 1000, namely the CpG site with largest loading value in HOXA5, TET1, and GABBR1 ranked 1059, 1142, and 1535 out of 450K respectively.
Concerning the fourth gene HOXA6, we performed univariate logistic regressions of the Down syndrome outcome on each of the 20 CpG sites located in the genetic region, and only identified one significant CpG site ($ p $-value of 0.018). In comparison, the other selected genes from the HOXA family members have more significant CpG sites (such as HOXA2 with 22, HOXA4 with 16, the borderline HOXA5 with 13), and smaller $ p $-values for the most associated CpG sites (HOXA2 0.0003, HOXA5 0.003). Furthermore, there is little evidence linking HOXA6 to the functions of glycans. Therefore, our proposed GLM-PO2PLS seems to better identify the CpG sites relevant to both DS and glycomics.

It is worth mentioning that we expect differences between our approach and the single-omic studies. The single-omic approaches did not	consider the presence of correlation between CpG sites and glycomics when modeling the association of CpG sites and Down syndrome. Therefore, some methylation-specific genes that are unrelated with glycomics can rank lower in the joint components in our analysis. Furthermore, in GLM-PO2PLS, we focus on the joint part and the omic-specific parts are not linked to the outcome variable, and hence the top genes mapped from the methylation-specific components are not necessarily associated with the outcome DS. In this regard, an extension of our model which also considers the omic-specific parts in the linear predictor for the outcome variable can provide further insights into the disease from omic-specific aspects.

We have shown evidence for association between the mapped gene set and Down syndrome. Nonetheless, The dedicated ``Down syndrome'' set in the DisGeNET database was not enriched in our gene set. One reason could be that very few studies have been conducted on DS with methylation data. Furthermore, common diseases and cancers are usually more frequently studied, resulting in possible publication bias in the database. We searched the genes identified by both our study and~\cite{Bacalini2015} (namely, RUNX1, DLL1, EBF4, HOXA2, HOXA4, HHIP, NCAM1, NAV1, PRDM8, KDM2B) in the DisGeNET database, and found none of these genes has  their highest association score (i.e., amount of evidence) with DS. For example, the RUNX1 gene had the highest association score of 0.8 with acute myeloid leukemia, and a score of only 0.1 with DS. 

When estimating a GLM-PO2PLS binary model, we rely on numerical integration. The computational complexity of the numerical estimation is $ \mathcal{O}(M^{2r}) $, with $ M $ nodes per dimension. 
In practice this means that the binary model can only include 1 joint component. 
A computationally feasible solution is to include only one pair of the joint components in the linear predictor for the binary outcome. Such a model might be suited for our Down syndrome analysis where only one pair of joint components was associated to the outcome. However, the assumption that only one pair of joint components is related to the outcome might not apply to other studies.
Therefore, a more efficient numerical integration strategy is needed. One strategy is to use adaptive quadrature. Although for a fixed number of nodes $ M $, the adaptive quadrature is computationally more complex than its non-adaptive counterpart we used, the adaptive variant needs a smaller $ M $ to reach an equally precise approximation, thus can be more efficient~\cite{Rabe-Hesketh2002,Rockwood2021a}. Another strategy is to decompose the $ 2r $-dimensional integration to $ r $ 2-dimensional integrations. This will reduce the computational complexity to $ \mathcal{O}(r\times M^{2}) $.

To calculate the $ p $-values for the tests in~\eqref{Testchi} and~\eqref{Testchi2}, we derived the asymptotic normality of the estimator for the parameters of GLM-PO2PLS with a normally distributed outcome. Asymptotic normality was proved by showing that the mapping (denote $ \tau $) from the parameter vector $ \theta $ to the moment structure as well as the discrepancy function with respect to the moments satisfy certain regularity conditions~\cite{Shapiro1986}. For the GLM-PO2PLS model with a binary outcome,  there is not an explicit mapping function $ \tau $, and it is difficult to parameterize the likelihood in terms of the moments. Therefore, while the $ p $-values for the binary model can be calculated assuming the asymptotic normality holds, it is unclear whether they are correct.
The derivation of asymptotic normality for the binary model is future work.

In this paper our aim was to use one model for all the data, and model the relationships between the omics simultaneously with their relationship with an outcome. While providing a holistic overview, for the binary outcome variable, the approach is computationally intensive. In our data analysis, it appeared that modeling the binary outcome as continuous provided similar information. Alternatively, a two-stage approach might be used. We recently proposed a two-stage PO2PLS approach~\cite{Gu2021a}, where we first constructed a few joint latent components that represent the two omics, then linked these latent components to the outcome variable using a linear regression model. In the implementation of two-stage PO2PLS to the DS dataset, the latent variables from the first stage were used as outcomes in several separate regression models in the second stage, thus the interpretation was different from a logistic regression model with DS as outcome. Alternatively, the latent variables can also be used as covariates in the second stage. However, the latent variables contain errors from the dimension reduction process. Ignoring these errors in the covariates can cause attenuated predicted probabilities in the logistic regression~\cite{Stefanski1985}. Therefore, to correctly model the outcome, the two-stage approach needs to be augmented with a measurement error model for the latent variables. Here more research is needed. 

Several extensions of GLM-PO2PLS might be relevant. For an outcome variable from other members of the exponential family (e.g., Poisson, gamma, etc.), the corresponding EM algorithm can be obtained by modifying the EM algorithm for the binary outcome by replacing $ p(z|\nu) $ in~\eqref{conpz} with the corresponding conditional probability mass/density function. 
Regarding the relationship between omics and outcome, the omic-specific latent variables are not included in the linear predictor for the outcome variable in GLM-PO2PLS.
As discussed above, linking the omic-specific parts to the outcome might provide further insights. Furthermore, since the omic-specific latent variable might also be predictive of the outcome, a model where all the latent variables are linked to the outcome can lead to improved outcome prediction performance in some studies. Extending GLM-PO2PLS to such a model will increase the computational complexity to $ \mathcal{O}(M^{2r+r_x+r_y}) $ for non-normal outcomes.
Another direction is to generalize the model to incorporate more than two omic datasets jointly with an outcome. Such an extension would require to specify the directions of the relationships among more than two sets of variables. A workaround might be to model a common set of latent variables for all sets of variables~\cite{Meng2016}. For some studies, the directions of the relationships are clear (e.g., among genetics, methylation, and glycomics), and specifying the direction in the model and allowing the joint latent variables for each set of variables to differ can improve model performance. However, the computation will also be intensive for a binary outcome.

To conclude, GLM-PO2PLS is a promising method to model an outcome with two omic datasets and as a base for further extensions.

\section*{Data availability statement}
The DNA methylation data used in this study are available at the NCBI Gene Expression Omnibus (GEO) (http://www.ncbi.nlm.nih.gov/geo/) under accession number GSE52588.

\section*{Disclosure statement}
The authors declared no potential conflicts of interest with respect to the research, authorship, and/or publication of this article.
\section*{Funding}
The authors were supported by the following financial support for the research, authorship, and/or publication of this article: Zhujie Gu was supported by the European Union’s Horizon 2020 research and innovation programme IMforFUTURE [grant number 721815]; and the EU/EFPIA Innovative Medicines Initiative 2 Joint Undertaking BigData@Heart [grant number 116074]. Said el Bouhaddani was supported by ERA-Net E-Rare JTC 2018 (MSA-omics) [40-44000-98-2006/ 90030376507].

\bibliographystyle{tfs}
\bibliography{library}

\begin{thebibliography}{10}
\providecommand{\MR}{\relax\unskip\space MR }
\providecommand{\url}[1]{\normalfont{#1}}
\providecommand{\urlprefix}{Available at }

\bibitem{Abramowitz1972}
M. Abramowitz and  {Irene Stegun}, \emph{{Numerical interpolation,
  differentiation, and integration}}, Handbook of mathematical functions
  (1972), pp. 877--925.
  \urlprefix\url{https://books.google.com/books/about/Handbook_of_Mathematical_Functions.html?id=V3ZQAAAAMAAJ}.

\bibitem{Armijo1966}
L. Armijo, \emph{{Minimization of functions having lipschitz continuous first
  partial derivatives}}, Pacific Journal of Mathematics 16 (1966), pp. 1--3.
  \urlprefix\url{https://projecteuclid.org/journals/pacific-journal-of-mathematics/volume-16/issue-1/Minimization-of-functions-having-Lipschitz-continuous-first-partial-derivatives/pjm/1102995080.full
  https://projecteuclid.org/journals/pacific-journal-of-mathematics/volum}.

\bibitem{Ashburner2000}
M. Ashburner, C.A. Ball, J.A. Blake, D. Botstein, H. Butler, J.M. Cherry, A.P.
  Davis, K. Dolinski, S.S. Dwight, J.T. Eppig, M.A. Harris, D.P. Hill, L.
  Issel-Tarver, A. Kasarskis, S. Lewis, J.C. Matese, J.E. Richardson, M.
  Ringwald, G.M. Rubin, and G. Sherlock, \emph{{Gene ontology: Tool for the
  unification of biology}} (2000). \urlprefix\url{/pmc/articles/PMC3037419/
  /pmc/articles/PMC3037419/?report=abstract
  https://www.ncbi.nlm.nih.gov/pmc/articles/PMC3037419/}.

\bibitem{Bacalini2015}
M.G. Bacalini, D. Gentilini, A. Boattini, E. Giampieri, C. Pirazzini, C.
  Giuliani, E. Fontanesi, M. Scurti, D. Remondini, M. Capri, G. Cocchi, A.
  Ghezzo, A.D. Rio, D. Luiselli, G. Vitale, D. Mari, G. Castellani, M. Fraga,
  A.M. {Di Blasio}, S. Salvioli, C. Franceschi, and P. Garagnani,
  \emph{{Identification of a DNA methylation signature in blood cells from
  persons with down syndrome}}, Aging 7 (2015), pp. 82--96.

\bibitem{Benjamini1995}
Y. Benjamini and Y. Hochberg, \emph{{Controlling the False Discovery Rate: A
  Practical and Powerful Approach to Multiple Testing}}, Journal of the Royal
  Statistical Society: Series B (Methodological) 57 (1995), pp. 289--300.

\bibitem{Borelli2015}
V. Borelli, V. Vanhooren, E. Lonardi, K.R. Reiding, M. Capri, C. Libert, P.
  Garagnani, S. Salvioli, C. Franceschi, and M. Wuhrer, \emph{{Plasma N-Glycome
  Signature of Down Syndrome}}, Journal of Proteome Research 14 (2015), pp.
  4232--4245.

\bibitem{Choi2019}
C. Choi, T. Kim, K.T. Chang, and K.T. Min, \emph{{DSCR1-mediated TET1 splicing
  regulates miR-124 expression to control adult hippocampal neurogenesis}}, The
  EMBO Journal 38 (2019), p. e101293.
  \urlprefix\url{https://onlinelibrary.wiley.com/doi/full/10.15252/embj.2018101293
  https://onlinelibrary.wiley.com/doi/abs/10.15252/embj.2018101293
  https://www.embopress.org/doi/abs/10.15252/embj.2018101293}.

\bibitem{Ciccarone2018}
F. Ciccarone, E. Valentini, M. Malavolta, M. Zampieri, M.G. Bacalini, R.
  Calabrese, T. Guastafierro, A. Reale, C. Franceschi, M. Capri, N. Breusing,
  T. Grune, M. {Morenoe Villanueva}, A. B{\"{u}}rkle, and P. Caiafa, \emph{{DNA
  Hydroxymethylation Levels Are Altered in Blood Cells from Down Syndrome
  Persons Enrolled in the MARK-AGE Project}}, Journals of Gerontology - Series
  A Biological Sciences and Medical Sciences 73 (2018), pp. 737--744.
  \urlprefix\url{/pmc/articles/PMC5946825/
  /pmc/articles/PMC5946825/?report=abstract
  https://www.ncbi.nlm.nih.gov/pmc/articles/PMC5946825/}.

\bibitem{Cindric2021}
A. Cindric, F. Vuckovic, V. Borelli, J. Juric, H. Deris, A. Murray, I. Alic, J.
  Groet, D. Petrovic, and S. Hamburg, \emph{{Accelerated biological aging in
  people with Down syndrome with full and segmental trisomy 21 begins in
  childhood as revealed by immunoglobulin G glycosylation}}, Research Square
  (2021), pp. 1--29.

\bibitem{TieJong1993}
S. de  Jong, \emph{{SIMPLS: An alternative approach to partial least squares
  regression}}, Chemometrics and Intelligent Laboratory Systems 18 (1993), pp.
  251--263.

\bibitem{Dempster1977}
A.P. Dempster, N.M. Laird, and D.B. Rubin, \emph{{ Maximum Likelihood from
  Incomplete Data Via the EM Algorithm }}, Journal of the Royal Statistical
  Society: Series B (Methodological) 39 (1977), pp. 1--22.

\bibitem{Bouhaddani2016a}
S. el  Bouhaddani, J. Houwing-Duistermaat, P. Salo, M. Perola, G. Jongbloed,
  and H.W. Uh, \emph{{Evaluation of O2PLS in Omics data integration}}, BMC
  Bioinformatics 17 (2016), p. S11.
  \urlprefix\url{https://bmcbioinformatics.biomedcentral.com/articles/10.1186/s12859-015-0854-z}.

\bibitem{ElBouhaddani2018}
S. el  Bouhaddani, H.W. Uh, C. Hayward, G. Jongbloed, and J.
  Houwing-Duistermaat, \emph{{Probabilistic partial least squares model:
  Identifiability, estimation and application}}, Journal of Multivariate
  Analysis 167 (2018), pp. 331--346.
  \urlprefix\url{https://doi.org/10.1016/j.jmva.2018.05.009}.

\bibitem{Bouhaddani2021}
S. el  Bouhaddani, H.W. Uh, G. Jongbloed, and J. Houwing-Duistermaat,
  \emph{{Statistical Integration of Heterogeneous Data with PO2PLS}}  (2021).
  \urlprefix\url{http://arxiv.org/abs/2103.13490}.

\bibitem{Franceschi2019}
C. Franceschi, P. Garagnani, N. Gensous, M.G. Bacalini, M. Conte, and S.
  Salvioli, \emph{{Accelerated bio-cognitive aging in Down syndrome: State of
  the art and possible deceleration strategies}} (2019).
  \urlprefix\url{https://onlinelibrary.wiley.com/doi/abs/10.1111/acel.12903}.

\bibitem{Gensous2020}
N. Gensous, M.G. Bacalini, C. Franceschi, and P. Garagnani, \emph{{Down
  syndrome, accelerated aging and immunosenescence}} (2020).
  \urlprefix\url{https://doi.org/10.1007/s00281-020-00804-1}.

\bibitem{Gu2021a}
Z. Gu, S. {El Bouhaddani}, J. Houwing-Duistermaat, and H.w. Uh,
  \emph{{Investigating the impact of Down syndrome on Methylation and Glycomics
  with two-stage PO2PLS}}, Theoretical Biology Forum  (2021), pp. 29--44.
  \urlprefix\url{http://digital.casalini.it/5213807}.

\bibitem{Haas2013}
M.A. Haas, D. Bell, A. Slender, E. Lana-Elola, S. Watson-Scales, E.M. Fisher,
  V.L. Tybulewicz, and F. Guillemot, \emph{{Alterations to dendritic spine
  morphology, but not dendrite patterning, of cortical projection neurons in
  Tc1 and Ts1Rhr mouse models of Down syndrome.}}, PloS one 8 (2013), p. 78561.
  \urlprefix\url{/pmc/articles/PMC3813676/
  /pmc/articles/PMC3813676/?report=abstract
  https://www.ncbi.nlm.nih.gov/pmc/articles/PMC3813676/}.

\bibitem{Hoerl2000}
A.E. Hoerl and R.W. Kennard, \emph{{Ridge Regression: Biased Estimation for
  Nonorthogonal Problems}}, Technometrics 42 (2000), p.~80.

\bibitem{Horvath2013}
S. Horvath, \emph{{DNA methylation age of human tissues and cell types}},
  Genome Biology 14 (2013), p. 115.
  \urlprefix\url{http://genomebiology.com//14/10/R115}.

\bibitem{Horvath2015}
S. Horvath, P. Garagnani, M.G. Bacalini, C. Pirazzini, S. Salvioli, D.
  Gentilini, A.M. {Di Blasio}, C. Giuliani, S. Tung, H.V. Vinters, and C.
  Franceschi, \emph{{Accelerated epigenetic aging in Down syndrome}}, Aging
  Cell 14 (2015), pp. 491--495.

\bibitem{Huo2018}
H.Q. Huo, Z.Y. Qu, F. Yuan, L. Ma, L. Yao, M. Xu, Y. Hu, J. Ji, A.
  Bhattacharyya, S.C. Zhang, and Y. Liu, \emph{{Modeling Down Syndrome with
  Patient iPSCs Reveals Cellular and Migration Deficits of GABAergic Neurons}},
  Stem Cell Reports 10 (2018), pp. 1251--1266.
  \urlprefix\url{https://doi.org/10.1016/j.stemcr.2018.02.001}.

\bibitem{Krautter2021}
F. Krautter and A.J. Iqbal, \emph{{Glycans and Glycan-Binding Proteins as
  Regulators and Potential Targets in Leukocyte Recruitment}} (2021).
  \urlprefix\url{www.frontiersin.org}.

\bibitem{Kristic2014}
J. Kri{\v{s}}ti{\'{c}}, F. Vu{\v{c}}kovi{\'{c}}, C. Menni, L. Klari{\'{c}}, T.
  Keser, I. Beceheli, M. Pu{\v{c}}i{\'{c}}-Bakovi{\'{c}}, M. Novokmet, M.
  Mangino, K. Thaqi, P. Rudan, N. Novokmet, J. {\v{S}}arac, S. Missoni, I.
  Kol{\v{c}}i{\'{c}}, O. Pola{\v{s}}ek, I. Rudan, H. Campbell, C. Hayward, Y.
  Aulchenko, A. Valdes, J.F. Wilson, O. Gornik, D. Primorac, V. Zoldo{\v{s}},
  T. Spector, and G. Lauc, \emph{{Glycans are a novel biomarker of
  chronological and biological ages}}, Journals of Gerontology - Series A
  Biological Sciences and Medical Sciences 69 (2014), pp. 779--789.
  \urlprefix\url{/pmc/articles/PMC4049143/
  /pmc/articles/PMC4049143/?report=abstract
  https://www.ncbi.nlm.nih.gov/pmc/articles/PMC4049143/}.

\bibitem{Li2017}
G. Li and S. Jung, \emph{{Incorporating covariates into integrated factor
  analysis of multi-view data}}, Biometrics 73 (2017), pp. 1433--1442.

\bibitem{Liu1994}
Q. Liu and D.A. Pierce, \emph{{A Note on Gauss-Hermite Quadrature}}, Biometrika
  81 (1994), p. 624.

\bibitem{Louis1982}
T.A. Louis, \emph{{Finding the Observed Information Matrix When Using the EM
  Algorithm}}, Journal of the Royal Statistical Society: Series B
  (Methodological) 44 (1982), pp. 226--233.

\bibitem{Meng2016}
C. Meng, O.A. Zeleznik, G.G. Thallinger, B. Kuster, A.M. Gholami, and A.C.
  Culhane, \emph{{Dimension reduction techniques for the integrative analysis
  of multi-omics data}}, Briefings in Bioinformatics 17 (2016), pp. 628--641.
  \urlprefix\url{https://pubmed.ncbi.nlm.nih.gov/26969681/}.

\bibitem{Min2018}
J.L. Min, G. Hemani, G.D. Smith, C. Relton, and M. Suderman, \emph{{Meffil:
  Efficient normalization and analysis of very large DNA methylation
  datasets}}, Bioinformatics 34 (2018), pp. 3983--3989.
  \urlprefix\url{https://pubmed.ncbi.nlm.nih.gov/29931280/}.

\bibitem{Mootha2003}
V.K. Mootha, C.M. Lindgren, K.F. Eriksson, A. Subramanian, S. Sihag, J. Lehar,
  P. Puigserver, E. Carlsson, M. Ridderstr{\aa}le, E. Laurila, N. Houstis, M.J.
  Daly, N. Patterson, J.P. Mesirov, T.R. Golub, P. Tamayo, B. Spiegelman, E.S.
  Lander, J.N. Hirschhorn, D. Altshuler, and L.C. Groop,
  \emph{{PGC-1$\alpha$-responsive genes involved in oxidative phosphorylation
  are coordinately downregulated in human diabetes}}, Nature Genetics 34
  (2003), pp. 267--273. \urlprefix\url{https://www.nature.com/articles/ng1180}.

\bibitem{Nelder1972a}
J.A. Nelder and R.W.M. Wedderburn, \emph{{Generalized Linear Models}}, Journal
  of the Royal Statistical Society. Series A (General) 135 (1972), p. 370.
  \urlprefix\url{https://www.jstor.org/stable/10.2307/2344614?origin=crossref}.

\bibitem{Nishiyama2021}
A. Nishiyama and M. Nakanishi, \emph{{Navigating the DNA methylation landscape
  of cancer}} (2021).
  \urlprefix\url{http://www.cell.com/article/S016895252100130X/fulltext
  http://www.cell.com/article/S016895252100130X/abstract
  https://www.cell.com/trends/genetics/abstract/S0168-9525(21)00130-X}.

\bibitem{Patterson2007}
D. Patterson, \emph{{Genetic mechanisms involved in the phenotype of down
  syndrome}} (2007). \urlprefix\url{https://pubmed.ncbi.nlm.nih.gov/17910086/
  https://pubmed.ncbi.nlm.nih.gov/17910086/?dopt=Abstract}.

\bibitem{Pinero2020}
J. Pi{\~{n}}ero, J.M. Ram{\'{i}}rez-Anguita, J. Sa{\"{u}}ch-Pitarch, F.
  Ronzano, E. Centeno, F. Sanz, and L.I. Furlong, \emph{{The DisGeNET knowledge
  platform for disease genomics: 2019 update}}, Nucleic Acids Research 48
  (2020), pp. D845--D855.
  \urlprefix\url{https://academic.oup.com/nar/article/48/D1/D845/5611674}.

\bibitem{Rabe-Hesketh2002}
S. Rabe-Hesketh, A. Skrondal, and A. Pickles, \emph{{Reliable Estimation of
  Generalized Linear Mixed Models using Adaptive Quadrature}}, The Stata
  Journal: Promoting communications on statistics and Stata 2 (2002), pp.
  1--21.

\bibitem{Rockwood2021a}
N.J. Rockwood, \emph{{Efficient Likelihood Estimation of Generalized Structural
  Equation Models with a Mix of Normal and Nonnormal Responses}}, Psychometrika
  86 (2021), pp. 642--667.
  \urlprefix\url{https://doi.org/10.1007/s11336-021-09770-5}.

\bibitem{Rodriguez-Girondo2018}
M. Rodr{\'{i}}guez-Girondo, P. Salo, T. Burzykowski, M. Perola, J.
  Houwing-Duistermaat, and B. Mertens, \emph{{Sequential double
  cross-validation for assessment of added predictive ability in
  high-dimensional omic applications}}, Annals of Applied Statistics 12 (2018),
  pp. 1655--1678.
  \urlprefix\url{https://projecteuclid.org/journals/annals-of-applied-statistics/volume-12/issue-3/Sequential-double-cross-validation-for-assessment-of-added-predictive-ability/10.1214/17-AOAS1125.full
  https://projecteuclid.org/journals/annals-of-applied-statistics/volume}.

\bibitem{Shapiro1986}
A. Shapiro, \emph{{Asymptotic theory of overparameterized structural models}},
  Journal of the American Statistical Association 81 (1986), pp. 142--149.

\bibitem{Sheikhpour2021}
M. Sheikhpour, M. Maleki, M. {Ebrahimi Vargoorani}, and V. Amiri, \emph{{A
  review of epigenetic changes in asthma: methylation and acetylation}} (2021).
  \urlprefix\url{https://clinicalepigeneticsjournal.biomedcentral.com/articles/10.1186/s13148-021-01049-x}.

\bibitem{Sobol2019}
M. Sobol, J. Klar, L. Laan, M. Shahsavani, J. Schuster, G. Anner{\'{e}}n, A.
  Konzer, J. Mi, J. Bergquist, J. Nordlund, J. Hoeber, M. Huss, A. Falk, and N.
  Dahl, \emph{{Transcriptome and Proteome Profiling of Neural Induced
  Pluripotent Stem Cells from Individuals with Down Syndrome Disclose Dynamic
  Dysregulations of Key Pathways and Cellular Functions}}, Molecular
  Neurobiology 56 (2019), pp. 7113--7127.
  \urlprefix\url{https://doi.org/10.1007/s12035-019-1585-3}.

\bibitem{Stagni2018}
F. Stagni, A. Giacomini, M. Emili, S. Guidi, and R. Bartesaghi,
  \emph{{Neurogenesis impairment: An early developmental defect in Down
  syndrome}} (2018).

\bibitem{Stefanski1985}
L.A. Stefanski and R.J. Carroll, \emph{{Covariate Measurement Error in Logistic
  Regression}} 13 (1985), pp. 1335--1351.

\bibitem{Storey2002}
J.D. Storey, \emph{{A direct approach to false discovery rates}}, Journal of
  the Royal Statistical Society. Series B: Statistical Methodology 64 (2002),
  pp. 479--498.

\bibitem{Subramanian2005}
A. Subramanian, P. Tamayo, V.K. Mootha, S. Mukherjee, B.L. Ebert, M.A.
  Gillette, A. Paulovich, S.L. Pomeroy, T.R. Golub, E.S. Lander, and J.P.
  Mesirov, \emph{{Gene set enrichment analysis: A knowledge-based approach for
  interpreting genome-wide expression profiles}}, Proceedings of the National
  Academy of Sciences of the United States of America 102 (2005), pp.
  15545--15550. \urlprefix\url{/pmc/articles/PMC1239896/?report=abstract
  https://www.ncbi.nlm.nih.gov/pmc/articles/PMC1239896/}.

\bibitem{Sugar2021}
S. Sug{\'{a}}r, G. T{\'{o}}th, F. Bugyi, K. V{\'{e}}key, K. Kar{\'{a}}szi, L.
  Drahos, and L. Turi{\'{a}}k, \emph{{Alterations in protein expression and
  site-specific N-glycosylation of prostate cancer tissues}}, Scientific
  Reports 11 (2021), pp. 1--12.
  \urlprefix\url{https://www.nature.com/articles/s41598-021-95417-5}.

\bibitem{Tabang2021}
D.N. Tabang, M. Ford, and L. Li, \emph{{Recent Advances in Mass
  Spectrometry-Based Glycomic and Glycoproteomic Studies of Pancreatic
  Diseases}} (2021).

\bibitem{Tissier2018b}
R. Tissier,  {Mar Rodr´ıguez-Girondo}, and J.J. Houwing-Duistermaat,
  \emph{{Integration of several omic sources in prediction models using
  network-based approaches}}, Biometrical Journal  (2018).

\bibitem{Trygg2003}
J. Trygg and S. Wold, \emph{{O2-PLS, a two-block (X-Y) latent variable
  regression (LVR) method with an integral OSC filter}}, in \emph{Journal of
  Chemometrics}, Vol.~17, jan. 2003, pp. 53--64.
  \urlprefix\url{http://doi.wiley.com/10.1002/cem.775}.

\bibitem{Uffelmann2021}
E. Uffelmann, Q.Q. Huang, N.S. Munung, J. de  Vries, Y. Okada, A.R. Martin,
  H.C. Martin, T. Lappalainen, and D. Posthuma, \emph{{Genome-wide association
  studies}}, Nature Reviews Methods Primers 2021 1:1 1 (2021), pp. 1--21.
  \urlprefix\url{https://www.nature.com/articles/s43586-021-00056-9}.

\bibitem{Uh2020}
H.W. Uh, L. Klaric, I. Ugrina, G. Lauc, A.K. Smilde, and J.J.
  Houwing-Duistermaat, \emph{{Choosing proper normalization is essential for
  discovery of sparse glycan biomarkers}}, Molecular Omics 16 (2020), pp.
  231--242.

\bibitem{Varki2017}
A. Varki, \emph{{Biological roles of glycans}}, Glycobiology 27 (2017), pp.
  3--49. \urlprefix\url{https://pubmed.ncbi.nlm.nih.gov/27558841/}.

\bibitem{Wahl2018}
A. Wahl, S. Kasela, E. Carnero-Montoro, M. van  Iterson, J. {\v{S}}tambuk, S.
  Sharma, E. van~den  Akker, L. Klaric, E. Benedetti, G. Razdorov, I.
  Trbojevi{\'{c}}-Akma{\v{c}}i{\'{c}}, F. Vu{\v{c}}kovi{\'{c}}, I. Ugrina, M.
  Beekman, J. Deelen, D. van  Heemst, B.T. Heijmans,  {B.I.O.S. Consortium}, M.
  Wuhrer, R. Plomp, T. Keser, M. {\v{S}}imurina, T. Pavi{\'{c}}, I. Gudelj, J.
  Kri{\v{s}}ti{\'{c}}, H. Grallert, S. Kunze, A. Peters, J.T. Bell, T.D.
  Spector, L. Milani, P.E. Slagboom, G. Lauc, and C. Gieger, \emph{{IgG
  glycosylation and DNA methylation are interconnected with smoking}},
  Biochimica et Biophysica Acta - General Subjects 1862 (2018), pp. 637--648.
  \urlprefix\url{https://www.sciencedirect.com/science/article/pii/S0304416517303410?dgcid=raven_sd_recommender_email}.

\bibitem{Watanabe2019}
K. Watanabe, S. Stringer, O. Frei, M. {Umi{\'{c}}evi{\'{c}} Mirkov}, C. de
  Leeuw, T.J. Polderman, S. van~der  Sluis, O.A. Andreassen, B.M. Neale, and D.
  Posthuma, \emph{{A global overview of pleiotropy and genetic architecture in
  complex traits}}, Nature Genetics 51 (2019), pp. 1339--1348.

\bibitem{WOLD1973}
H. WOLD, \emph{{Nonlinear Iterative Partial Least Squares (NIPALS) Modelling:
  Some Current Developments}}, in \emph{Multivariate Analysis–III},  1973,
  pp. 383--407.
  \urlprefix\url{https://www.sciencedirect.com/science/article/pii/B9780124266537500326}.

\bibitem{Yun2021}
Y. Yun, Y. Zhang, C. Zhang, L. Huang, S. Tan, P. Wang, C.
  Vilari{\~{n}}o-G{\'{u}}ell, W. Song, and X. Sun, \emph{{Regulator of
  calcineurin 1 is a novel RNA-binding protein to regulate neuronal
  apoptosis}}, Molecular Psychiatry 26 (2021), pp. 1361--1375.
  \urlprefix\url{https://pubmed.ncbi.nlm.nih.gov/31451750/}.

\bibitem{Zhang2016}
X. Zhang and Y. Wang, \emph{{Glycosylation Quality Control by the Golgi
  Structure}} (2016). \urlprefix\url{/pmc/articles/PMC4983240/
  /pmc/articles/PMC4983240/?report=abstract
  https://www.ncbi.nlm.nih.gov/pmc/articles/PMC4983240/}.

\bibitem{Zou2005}
H. Zou and T. Hastie, \emph{{Erratum: Regularization and variable selection via
  the elastic net (Journal of the Royal Statistical Society. Series B:
  Statistical Methodology (2005) 67 (301-320))}} (2005).

\end{thebibliography}

\end{document}